\documentclass[pra,twocolumn,amsmath,amssymb,floatfix,reprint,footinbib,superscriptaddress,longbibliography,showkeys]{revtex4-1}
\usepackage{dcolumn}
\usepackage{graphicx}
\usepackage{mathrsfs}
\usepackage{mdwlist}
\usepackage{subfigure}
\usepackage{booktabs}
\usepackage{amsmath}
\usepackage{extarrows}
\usepackage{dsfont}
\usepackage{amstext}
\usepackage{amssymb}
\usepackage{amsbsy}
\usepackage{bbm}
\usepackage{bm}
\usepackage{amsthm}
\usepackage{graphicx}
\usepackage{textcomp}
\usepackage{multirow}
\usepackage{color}
\usepackage{diagbox}
\usepackage[colorlinks,citecolor=blue]{hyperref}
\definecolor{Dgreen}{RGB}{0, 100, 0}
\usepackage{url}
\usepackage[colorlinks]{hyperref}
\hypersetup{%
	plainpages=true,
	breaklinks=true,  
	hypertexnames=false,  
	pageanchor=true,
	colorlinks=true,
	linkcolor={blue},
	citecolor={blue},
	urlcolor={blue},
	anchorcolor={black}
}

\begin{document}

\title{Robust single-qubit gates by composite pulses in three-level systems}

\author{Zhi-Cheng Shi}
\affiliation{Fujian Key Laboratory of Quantum Information and Quantum Optics (Fuzhou University), Fuzhou 350108, China}
\affiliation{Department of Physics, Fuzhou University, Fuzhou 350108, China}

\author{Hai-Ning Wu}
\affiliation{Fujian Key Laboratory of Quantum Information and Quantum Optics (Fuzhou University), Fuzhou 350108, China}
\affiliation{Department of Physics, Fuzhou University, Fuzhou 350108, China}

\author{Li-Tuo Shen}
\affiliation{Fujian Key Laboratory of Quantum Information and Quantum Optics (Fuzhou University), Fuzhou 350108, China}
\affiliation{Department of Physics, Fuzhou University, Fuzhou 350108, China}

\author{Jie Song}
\affiliation{Department of Physics, Harbin Institute of Technology, Harbin 150001, China}

\author{Yan Xia}\thanks{xia-208@163.com}
\affiliation{Fujian Key Laboratory of Quantum Information and Quantum Optics (Fuzhou University), Fuzhou 350108, China}
\affiliation{Department of Physics, Fuzhou University, Fuzhou 350108, China}

\author{X. X. Yi} \thanks{yixx@nenu.edu.cn}
\affiliation{Center for Quantum Sciences and School of Physics, Northeast Normal University, Changchun 130024, China}

\author{Shi-Biao Zheng}
\affiliation{Fujian Key Laboratory of Quantum Information and Quantum Optics (Fuzhou University), Fuzhou 350108, China}
\affiliation{Department of Physics, Fuzhou University, Fuzhou 350108, China}

\begin{abstract}
Composite pulses are an efficient tool for robust quantum control.
In this work, we derive the form of the composite pulse sequence to implement robust single-qubit gates in a three-level system, where two low-energy levels act as a qubit. The composite pulses can efficiently cancel the systematic errors up to a certain order. We find that the three-pulse sequence cannot completely eliminate the first order of systematic errors, but still availably makes the fidelity resistant to variations in a specific direction. When employing more pulses in the sequence ($N>3$), the fidelity can be insensitive to the variations in all directions and the robustness region becomes much wider.
Finally we demonstrate the applications of composite pulses in quantum information processing, e.g., robust quantum
information transfer between two qubits.

\end{abstract}

\maketitle

\section{Introduction}

Manipulating quantum states in a robust fashion is the key factor for quantum information processing \cite{Nielsen00}, and has attracted a lot of attention in recent decades. One of the well-known techniques in this field is the adiabatic passage \cite{physchem.52.1.763,RevModPhys.79.53,RevModPhys.89.015006}, which is insensitive to the errors induced by parameter imperfections.
Nevertheless, the drawbacks of slow evolution speed and not very high fidelity hinder the sustainable development of the adiabatic passage in quantum information processing.
To overcome these drawbacks, some other control techniques are proposed, such as
the smooth analytical pulse \cite{daems13,Barnes2015,PhysRevA.96.022309,Zeng2018,Gngrd2019},
invariant-based inverse engineering \cite{PhysRevA.83.062116,Ruschhaupt2012,PhysRevA.88.033406,
PhysRevA.100.023415,PhysRevA.95.022332,Levy2018,PhysRevA.97.062317,PhysRevA.90.063425,kang20},
robust optimal control \cite{PhysRevA.49.2241,PhysRevA.66.063405,PhysRevA.70.063412,PhysRevA.81.032312,PhysRevA.86.012317,PhysRevA.89.022341,
PhysRevA.90.062309,PhysRevLett.115.190801,Glaser2015,PhysRevA.82.013415,Dong2016,PhysRevA.95.063403,PhysRevA.99.042327,PhysRevA.102.043707},
composite pulses (CPs) \cite{Abraham61,Slichter90,Freeman97}, etc.
Among them, CPs are a powerful tool for precise quantum state manipulations, since CPs possess the merits of both ultrahigh fidelity and robustness against systematic errors.

The basic idea of CPs is to make up for the deviations of system parameters by constructing a sequence of pulses.
The control variables, such as the time durations \cite{Wang2012,PhysRevA.89.022310,PhysRevLett.110.140502,PhysRevA.97.042324}, the detunings \cite{PhysRevA.99.013424,PhysRevA.100.032333}, or the phases \cite{PhysRevLett.113.043001,PhysRevA.84.065404,PhysRevA.88.063410,PhysRevA.99.013402,PhysRevA.101.012321,PhysRevA.101.013827}, are meticulously designed to compensate for the systematic errors to any desired order \cite{Tomita2010,PhysRevA.90.033608,PhysRevA.90.040301,PhysRevA.93.032340,Ivanov2013,PhysRevLett.118.150502}. As a result, quantum manipulations would increase the robustness with respect to the systematic errors.
Until now, CP studies \cite{PhysRevA.70.052318,Wang2012,PhysRevLett.106.233001,PhysRevLett.110.140502,PhysRevA.89.022310,PhysRevA.90.012316, PhysRevA.87.052317,PhysRevA.88.063410,PhysRevA.92.042304,PhysRevLett.113.043001,PhysRevA.99.013402,PhysRevA.84.065404,PhysRevA.93.023830,
PhysRevA.97.042324,PhysRevA.99.013424,PhysRevA.100.032333,PhysRevLett.118.133202,PhysRevA.100.023410,PhysRevA.101.012321,
PhysRevA.102.013105,PhysRevA.101.013827} concentrated mainly on the simplest two-level systems, and were really successful in eliminating all kinds of errors caused by the inhomogeneities of parameters.
However, little attention has been paid to the three-level systems \cite{PhysRevA.84.063413,PhysRevA.98.043414,Greener2018,PhysRevResearch.2.043194,PhysRevResearch.2.043235}, since the construction of CPs requires one to manage the complicated multilevel dynamics.

As is well known, quantum systems are hardly isolated from the environment, which gives rise to the decoherence effect \cite{Breuer2006}. In two-level systems, the coherence of a qubit would be gradually diminished during quantum operations because the high-energy level unavoidably interacts with the environment.
There are two ways to resist the decoherence. The first way is to quickly accomplish the qubit operations before the coherence completely vanishes. Since the magnitude of the coupling strength in different quantum systems is usually determinate at current technologies (e.g., about the order of megahertz in atomic systems \cite{Omran2019,Madjarov2020}), the manipulation time is hard to be significantly shortened (about the order of microseconds in atomic systems). Another way is to increase the coherence time by the $\Lambda$-type three-level system. More specifically, one can encode the qubit into two low-energy levels, and the transition between two low-energy levels is indirectly achieved through the medium of a high-energy level.
On the other hand, the three-level systems are essential for several quantum operations. For instance, to implement the transition from the ground state to the Rydberg state, we must resort to the intermediate state \cite{PhysRevLett.114.113002}. Additionally, some complicated quantum systems can be reduced to the three-level physical models under specified conditions.
Therefore, it is very necessary to study the issue of implementing robust quantum control by designing a sequence of pulses in three-level systems.

In this work, we consider a three-level system in which two low-energy levels, acting as a qubit, cannot be directly coupled each other, so they require a high-energy level to construct the indirect coupling. We present a general theoretical method to implement robust single-qubit gates by CPs. Those pulses effectively compensate for the systematic errors caused by the variations in the external fields.
Our method is based on the Taylor expansion of the actual evolution operator, and the results demonstrate that the higher order of the systematic errors can be eliminated by CPs with the increasing of the number of pulses. Therefore, CPs would exhibit a prominently robust performance for the variations.
As an example of relevant applications, we finally show how to implement robust quantum information transfer between two qubits by CPs.

The rest of the paper is organized as follows. In Sec.~\ref{II}, we present the problem that the implementation of universal single-qubit gates by resonant pulse is severely affected by the variations, and demonstrate how to effectively restrain the influence of variations by CPs. In Sec.~\ref{III}, we illustrate that the $N$-pulse sequence ($N=3,5,7$) can be used to implement different robust single-qubit gates. In Sec.~\ref{IV}, we show the applications of CPs in performing robust quantum computations, such as the quantum information transfer between two qubits. A conclusion is given in Sec.~\ref{V}.

\section{General theory for robust control of single-qubit gates}  \label{II}

We consider the physical model in which a $\Lambda$-type three-level system has two low-energy levels ($|1\rangle$ and $|2\rangle$) and a high-energy level ($|3\rangle$). The transition $|1\rangle\leftrightarrow|3\rangle$ ($|2\rangle\leftrightarrow|3\rangle$) is resonantly driven by the external field with the coupling strength $\Omega_1$ ($\Omega_2$) and the phase $\alpha$ ($\beta$). In the presence of the variations in the external fields, the system Hamiltonian can be written as ($\hbar=1$)
\begin{equation} \label{0}
H\!=\!(1+\delta_1)\Omega_1\!\exp({i\alpha})|1\rangle\!\langle3|\!+ \!(1+\delta_2)\Omega_2\!\exp({i\beta})|2\rangle\!\langle3|+\mathrm{H.c.},
\end{equation}
where the variations $\delta_1$ and $\delta_2$ are random unknown constants. Here, we consider that the variations are caused by the inhomogeneities of the external fields, which is common in quantum systems.
For instance, when the atoms are resonantly driven by the laser pulses, the coupling strengths are distinct in different positions, because the spatial distribution of laser fields is not uniform.
Due to the imperfect knowledge of the spatial position of atoms, there exist variations in the coupling strength.
Besides, $\delta_1$ and $\delta_2$ in Eq.~(\ref{0}) can also represent the variations in the interaction time.
Note that two low-energy levels $|1\rangle$ and $|2\rangle$ are viewed as a qubit, and they cannot be directly coupled with each other. Thus, a high-energy level $|3\rangle$ is required to construct the transition between two low-energy levels.
It is worth mentioning that we do not focus on the specific physical system for the Hamiltonian in Eq.~(\ref{0}), since this physical model can be easily implemented in various physical systems, such as atomic systems \cite{scully97}, trapped ions \cite{RevModPhys.75.281}, superconducting systems \cite{RevModPhys.85.623}, etc.

\begin{figure}[t]
	\centering
	\includegraphics[scale=0.062]{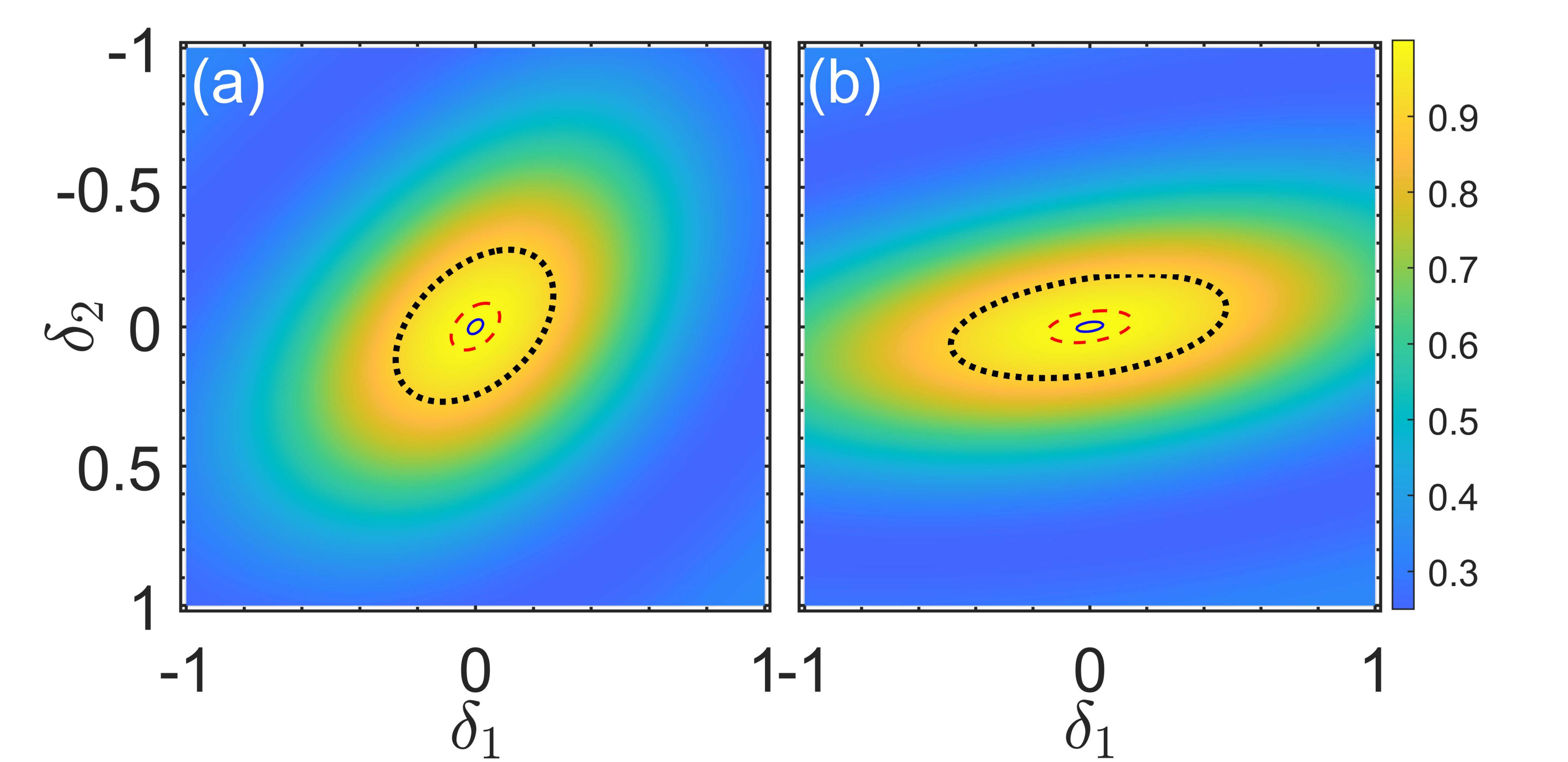}
	\caption{Fidelity $F$ vs the variations $\delta_1$ and $\delta_2$ by the resonant pulse, where $\alpha=\beta=0$. (a) $\Omega_2=\Omega_1$. (b) $\Omega_2=2\Omega_1$. We obtain distinct single-qubit gates in (a) and (b). Henceforth, the solid blue curve, the dashed red curve, and the dotted black curve correspond to $F=0.999$, 0.99, and 0.9, respectively. The results show that the universal single-qubit gates are sensitive to the variations $\delta_1$ and $\delta_2$, since the area encircled by the solid blue (dashed red) curve is quite small.}  \label{fig:00}
\end{figure}

First of all, in the absence of the variations (i.e., $\delta_1=0$ and $\delta_2=0$), one can easily obtain a universal single-qubit gate by choosing the evolution time $T=\pi/\sqrt{\Omega_1^2+\Omega_2^2}$, and the evolution operator in the basis $\{|1\rangle,|2\rangle,|3\rangle\}$ reads
\begin{eqnarray}  \label{1}
\mathbb{U}_{0}=\left[
                \begin{array}{ccc}
                 \cos2\chi & \exp({i\gamma})\sin2\chi & 0 \\
                 \exp({-i\gamma})\sin2\chi & -\cos2\chi & 0 \\
                 0  &  0 &  \exp({i\Upsilon})
                \end{array}
              \right],
\end{eqnarray}
where $\tan\chi=\Omega_2/\Omega_1$ and $\gamma=\alpha-\beta$. The values of $\chi$ and $\gamma$ are determined by the type of single-qubit gate we desire to obtain. The phase $\Upsilon$ is irrelevant to the qubit in the present model, thus we ignore it hereafter. Note that the fidelity of the single-qubit gate would sharply deteriorate when the external fields are deviated. This can be seen by Taylor expansion of the actual evolution operator $\mathbb{U}_a$ in the vicinity of $\delta_1=0$ and $\delta_2=0$,
\begin{equation}
\mathbb{U}_a=\mathbb{U}_{0}+\mathbb{U}_{e},
\end{equation}
where $\mathbb{U}_{0}$ is represented by Eq.~(\ref{1}), and $\mathbb{U}_{e}$ denotes the error operator induced by the variations.
The form of the error operator can be written as
\begin{equation}  \label{4}
\mathbb{U}_e=\mathbb{U}_{1}\delta_1
+\mathbb{U}_{2}\delta_2+{O}(\delta_1\delta_2,\delta_1^2,\delta_2^2),
\end{equation}
where
\begin{eqnarray}
\mathbb{U}_{1}&=&\left[
                \begin{array}{ccc}
                 g_1 & -g_2 & g_3 \\
                 -g_2^* & -g_1 & g_4 \\
                -g_3^* & -g_4^* & 0
                \end{array}
              \right],
~~\mathbb{U}_{2}=\left[
                \begin{array}{ccc}
                 -g_1 & g_2 & g_5 \\
                 g_2^* & g_1 & g_6 \\
                 -g_5^* & -g_6^* & 0
                \end{array}
              \right].   \nonumber
\end{eqnarray}
The elements are $g_1=\sin^{2}2\chi$, $g_2=\frac{1}{2}\exp({i\gamma})\sin4\chi$, $g_3=-i\exp({i\alpha})\pi\cos^{3}\chi$, $g_4=-i\exp({i\beta})\pi\cos^{2}\chi\sin\chi$, $g_5=-i\exp({i\alpha})\pi\cos\chi\sin^{2}\chi$, and $g_6=-i\exp({i\beta})\pi\sin^{3}\chi$.
One can observe from Eq.~(\ref{4}) that the variations $\delta_1$ and $\delta_2$ would lead to two types of errors. The first type is the qubit error which is caused by the imprecise coupling strength, reflected in the elements $g_{1}$ and $g_{2}$.
The second type is the leakage error, i.e., the population leakage from the low-energy levels to the high-energy level, reflected in the elements $g_{3}$, $g_{4}$, $g_{5}$, and $g_{6}$.
In Fig.~\ref{fig:00}, we plot the fidelity of the single-qubit gate as a function of the variations $\delta_1$ and $\delta_2$, where the fidelity is defined by \cite{Pedersen2007,Ghosh10}
\begin{eqnarray}
F=\frac{\mathrm{Tr}(\mathbb{U}_a\mathbb{U}^{\dag}_a)+|\mathrm{Tr}(\mathbb{U}^{\dag}_{0}\mathbb{U}_a)|^2}{d(d+1)},  ~~~d=3.
\end{eqnarray}
$\mathbb{U}_0$ is the desired single-qubit gate and $\mathbb{U}_a$ is the actual evolution operator.
The results verify that the fidelity of the single-qubit gate is extremely susceptible to the variations by the resonant pulse.

The target of this work is to design a composite $N$-pulse sequence with different phases to implement the robust universal single-qubit gate. In order to reduce the complexity of experimental operations, we demand that the coupling strengths remain unchanged during the whole sequence and the pulse intervals are assumed to be equidistant.
For the $n$th pulse, the Hamiltonian can be written as ($n=1, \dots, N$)
\begin{equation}
H_{n}\!=\!(1\!+\delta_1)\Omega_1\!\exp({i\alpha_{n}})\!|1\rangle\!\langle3|\!+\!(1\!+\delta_2)\Omega_2\!\exp({i\beta_{n}}) \!|2\rangle\!\langle3|\!+\!\mathrm{H.c.}
\end{equation}
Similarly, by Taylor expansion, the actual evolution operator $\mathbb{U}_a$ for these composite pulses can be expressed as
\begin{eqnarray}
\mathbb{U}_a&=&\exp({-iH_NT})\cdots\exp({-iH_1T})=\mathbb{U}_{0}+\mathbb{U}_{e},
\end{eqnarray}
where $\mathbb{U}_0$ is the desired single-qubit gate (in the absence of variations), and the general expression is given by
\begin{widetext}
\begin{equation}
\mathbb{U}_0\!=\!\!\left[
                \begin{array}{ccc}
                 \!\!\exp({i\phi_N})\!\cos2\theta_{N} & \exp({-i\varphi_N})\!\sin2\theta_{N} & 0 \\
                 \!\!-\exp({i\varphi_N})\!\sin2\theta_{N} & \exp({-i\phi_N})\!\cos2\theta_{N} & 0 \\
                 0  &  0 &  \!\!\!\!\!\exp({i\Upsilon_N})
                \end{array}
              \right].
\end{equation}
The coefficients $\phi_{n}$, $\varphi_{n}$, and $\theta_{n}$ ($n=2,\dots,N$) satisfy the following recursive relations:
\begin{eqnarray}
\phi_{n}&=&\arctan\frac{\cos2\theta_{n-1}\cos\phi_{n-1}\cos2\chi-\cos(\gamma_n+\varphi_{n-1})\sin2\theta_{n-1}\sin2\chi}{ \cos2\theta_{n-1}\sin\phi_{n-1}\cos2\chi-\sin(\gamma_n+\varphi_{n-1})\sin2\theta_{n-1}\sin2\chi},  \cr\cr
\varphi_{n}&=&\arctan\frac{\sin2\theta_{n-1}\cos\varphi_{n-1}\cos2\chi+\cos(\gamma_n-\varphi_{n-1})\cos2\theta_{n-1}\sin2\chi}{ \sin2\theta_{n-1}\sin\varphi_{n-1}\cos2\chi-\sin(\gamma_n-\varphi_{n-1})\cos2\theta_{n-1}\sin2\chi},    \cr\cr
\theta_{n}&=&\frac{1}{2}\arccos\sqrt{\cos^{2}2\theta_{n-1}\cos^{2}2\chi+\sin^{2}2\theta_{n-1}\sin^{2}2\chi- \frac{1}{2}\cos(\gamma_n-\phi_{n-1}+\varphi_{n-1})\sin4\theta_{n-1}\sin4\chi},
\end{eqnarray}
\end{widetext}
where $\tan\chi=\Omega_2/\Omega_1$, $\gamma_n=\alpha_n-\beta_n$, $\phi_{1}=\pi/2$, $\varphi_{1}=-\gamma_1-\pi/2$, and $\theta_{1}=\chi$.
$\mathbb{U}_{e}$ is the error operator induced by the variations $\delta_1$ and $\delta_2$, and the form is
\begin{equation}  \label{8}
\mathbb{U}_e=\mathbb{U}'_{1}\delta_1
+\mathbb{U}'_{2}\delta_2+{O}(\delta_1\delta_2,\delta_1^2,\delta_2^2),
\end{equation}
where
\begin{eqnarray}
\mathbb{U}'_{1}&=&\left[
                \begin{array}{ccc}
                 g'_1 & -g'_2 & g'_3 \\
                 -g^{\prime*}_2 & -g'_1 & g'_4 \\
                -g^{\prime*}_3 & -g^{\prime*}_4 & 0
                \end{array}
              \right],
~~\mathbb{U}'_{2}=\left[
                \begin{array}{ccc}
                 -g'_1 & g'_2 & g'_5 \\
                 g^{\prime*}_2 & g'_1 & g'_6 \\
                 -g^{\prime*}_5 & -g^{\prime*}_6 & 0
                \end{array}
              \right].   \nonumber
\end{eqnarray}
In order to distinguish the error operator in Eq.~(\ref{4}), we add the symbol ``$\prime$'' to the elements in Eq.~(\ref{8}).
By designing the values of phases $\alpha_n$ and $\beta_n$, we demand some (all) elements of $\mathbb{U}'_{1}$ and $\mathbb{U}'_{2}$ vanish so that the universal single-qubit gate is robust against the variations $\delta_1$ and $\delta_2$.
Remarkably, if we eliminate the elements of higher-order terms (e.g., $\delta_1\delta_2$, $\delta_1^2$, $\delta_2^2$, $\dots$) in the error operator $\mathbb{U}_e$, the fidelity of the single-qubit gate would be more robust against the variations.

\section{Examples}   \label{III}

In this section, we consider an odd number of symmetric composite pulses. The rationale behind this is that we can properly reduce the constraint conditions in the error operator $\mathbb{U}_{e}$ [e.g., cf. the relation between $g_3'$ and $g_5'$ in Eq.~(\ref{6c})].
For briefness, we take $N=3$ as the first example, and the form of the actual evolution operation reads
\begin{eqnarray}
\mathbb{U}_a=\exp({-iH_1T})\exp({-iH_2T})\exp({-iH_1T}). \nonumber
\end{eqnarray}
After some algebraic calculations, the expressions of the elements $g_{j}'$ ($j=1,\dots,6$) can be written as
\begin{widetext}
\begin{subequations}  \label{6}
\begin{eqnarray}     \label{6a}
\!\!\!\!\!\!\!\!\!\!\!g_1'&=&\frac{4\Omega_1^2\Omega_2^2}{(\Omega_1^2+\Omega_2^2)^4} \left[5\Omega_1^4-14\Omega_1^2\Omega_2^2+5\Omega_2^4 -4(\Omega_1^4-4\Omega_1^2\Omega_2^2+\Omega_2^4)\cos(\alpha_{12}-\beta_{12})\right],  \\ \label{6b}
\!\!\!\!\!\!\!\!\!\!\!g_2'&=&\frac{2\exp({i\gamma_1})\Omega_1\Omega_2(\Omega_1^2-\Omega_2^2)} {(\Omega_1^2+\Omega_2^2)^4}\left\{ 12\Omega_1^2\Omega_2^2\exp({i\alpha_{12}-i\beta_{12}})+ (\Omega_1^4-10\Omega_1^2\Omega_2^2+\Omega_2^4)\Big[2-\exp({i\beta_{12}-i\alpha_{12}})\Big]\right\},  \\ \label{6c}
\!\!\!\!\!\!\!\!\!\!\!g_3'&=&\frac{\Omega_1^2}{\Omega_2^2}g_5'\!=\!\frac{i\exp({i\alpha_1})\pi\Omega_1^3} {(\Omega_1^2+\Omega_2^2)^{7/2}} \!\!\left[ 2\Omega_2^2\exp({-i\beta_{12}})+(\Omega_1^2-\Omega_2^2)\exp({-i\alpha_{12}})\right] \!\! \left[2\Omega_1^2\exp({i\alpha_{12}})+2\Omega_2^2\exp({i\beta_{12}})+\Omega_1^2+\Omega_2^2\right]\!,  \\  \label{6d}
\!\!\!\!\!\!\!\!\!\!\!g_4'&=&\frac{\Omega_2^2}{\Omega_1^2}g_6'\!=\! \frac{i\exp({i\beta_1})\pi\Omega_1^2\Omega_2}{(\Omega_1^2+\Omega_2^2)^{7/2}}\!\! \left[ 2\Omega_1^2\exp({-i\alpha_{12}})\!-\!(\Omega_1^2\!-\!\Omega_2^2)\exp({-i\beta_{12}})\right] \!\! \left[2\Omega_1^2\exp({i\alpha_{12}})\!+\!2\Omega_2^2\exp({i\beta_{12}})\!+\!\Omega_1^2\!+\!\Omega_2^2\right]\!,
\end{eqnarray}
\end{subequations}
\end{widetext}
where the phase difference $\alpha_{12}=\alpha_1-\alpha_2$ and $\beta_{12}=\beta_1-\beta_2$.
Thus, in order to eliminate all first-order terms of the error operator $\mathbb{U}_e$,
the equations $g_1'=g_2'=g_3'=g_4'=0$ should be satisfied.
It is easily found that $g_2'=0$ if and only if $\Omega_1=\Omega_2$. For arbitrary values of $\Omega_1$ and $\Omega_2$, the solution of the equation $g_1'=0$ is given by
\begin{eqnarray}  \label{7}
\cos(\alpha_{12}-\beta_{12})=\frac{5\Omega_1^4-14\Omega_1^2\Omega_2^2+5\Omega_2^4} {4(\Omega_1^4-4\Omega_1^2\Omega_2^2+\Omega_2^4)},
\end{eqnarray}
and the solutions of the equations $g_3'=g_4'=0$ are
\begin{equation}  \label{8a}
\alpha_{12}\!=\!\arccos\frac{3\Omega_2^2-5\Omega_1^2}{4\Omega_1^2}, ~~\beta_{12}\!=\!-\arccos\frac{3\Omega_1^2-5\Omega_2^2}{4\Omega_2^2}.
\end{equation}

Note that there are only two variables (i.e., $\alpha_{12}$ and $\beta_{12}$) in the three-pulse sequence.
So, it is hard to simultaneously satisfy all equations by these two variables.
This means that we cannot completely eliminate all first-order terms of the error operator $\mathbb{U}_{e}$ by merely adopting the three-pulse sequence.
However, the three-pulse sequence can still effectively improve the fidelity of the universal single-qubit gate, since we can design the phases $\alpha_m$ and $\beta_m$ ($m=1,2$) to eliminate one kind of error, either the leakage error or the qubit error.
To be specific, if we set $\Omega_2=\Omega_1$ and the phases satisfy Eq.~(\ref{7}), the qubit error can be effectively removed. If the phases satisfy Eq.~(\ref{8a}), the leakage error can be effectively removed.

\begin{figure}[t]
	\centering
	\includegraphics[scale=0.061]{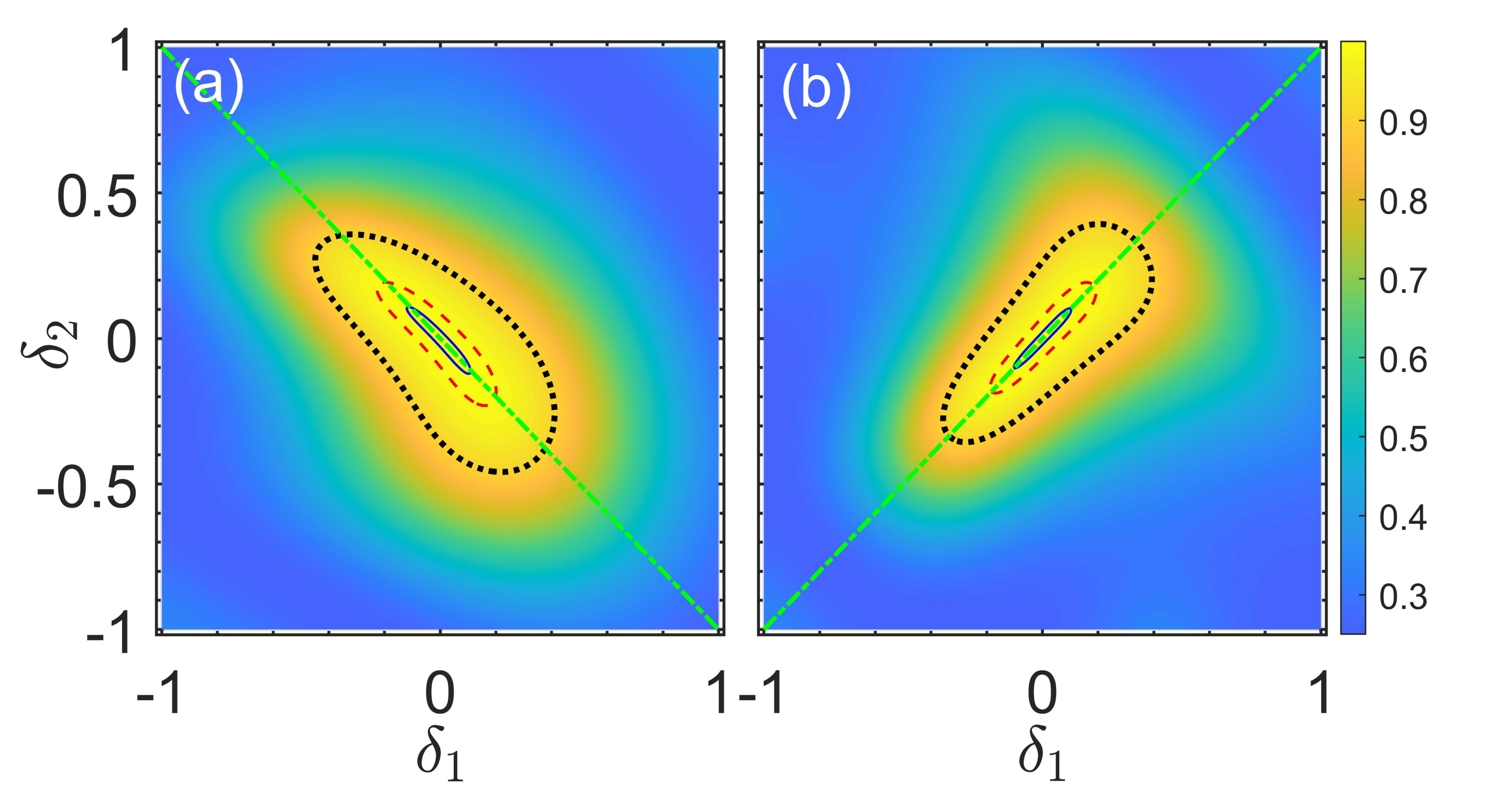}
	\caption{Fidelity $F$ vs the variations $\delta_1$ and $\delta_2$ in the three-pulse sequence, where $\Omega_2=\Omega_1$, and the phases $\alpha_{12}$ and $\beta_{12}$ are chosen to satisfy (a) Eq.~(\ref{7}) and (b) Eq.~(\ref{8a}), respectively. Compared with Fig.~\ref{fig:00}, the NOT gate obtained by the three-pulse sequence is less sensitive to the variations in a specific direction.}  \label{fig:01}
\end{figure}

In Fig.~\ref{fig:01}, we plot the fidelity $F$ as a function of the variations $\delta_1$ and $\delta_2$ in the three-pulse sequence, where $\Omega_2=\Omega_1$. In this situation, we obtain the NOT gate, and the form reads
\begin{eqnarray}
\mathbb{U}_{0}=\left[
                \begin{array}{ccc}
                 0 & \exp({i\frac{\pi}{3}}) & 0 \\
                 \exp({-i\frac{\pi}{3}}) & 0 & 0 \\
                 0  &  0 &  1
                \end{array}
              \right].
\end{eqnarray}
As can be seen from Fig.~\ref{fig:01}, the robustness behaviors have a significant difference for distinct phases. When the phases $\alpha_{12}$ and $\beta_{12}$ satisfy Eq.~(\ref{7}), the fidelity is robust against the variations around $\delta_1=-\delta_2$ [the green (light gray) line in Fig.~\ref{fig:01}(a)].
When the phases $\alpha_{12}$ and $\beta_{12}$ satisfy Eq.~(\ref{8a}), the fidelity is robust against the variations around $\delta_1=\delta_2$ [the green (light gray) line in Fig.~\ref{fig:01}(b)].
Apparently, these results are also different from the resonant pulse case, where the fidelity in Fig.~\ref{fig:00} is sensitive to the variations along all directions.
Note that there are four adjustable phases (i.e., $\alpha_{1}$, $\alpha_{2}$, $\beta_{1}$, and $\beta_{2}$) in the three-pulse sequence. So the values of $\alpha_{1}$ and $\beta_{1}$ can be employed to implement different single-qubit gates, and do not affect the solutions of $\alpha_{12}$ and $\beta_{12}$.

\begin{figure}[t]
	\centering
	\includegraphics[scale=0.061]{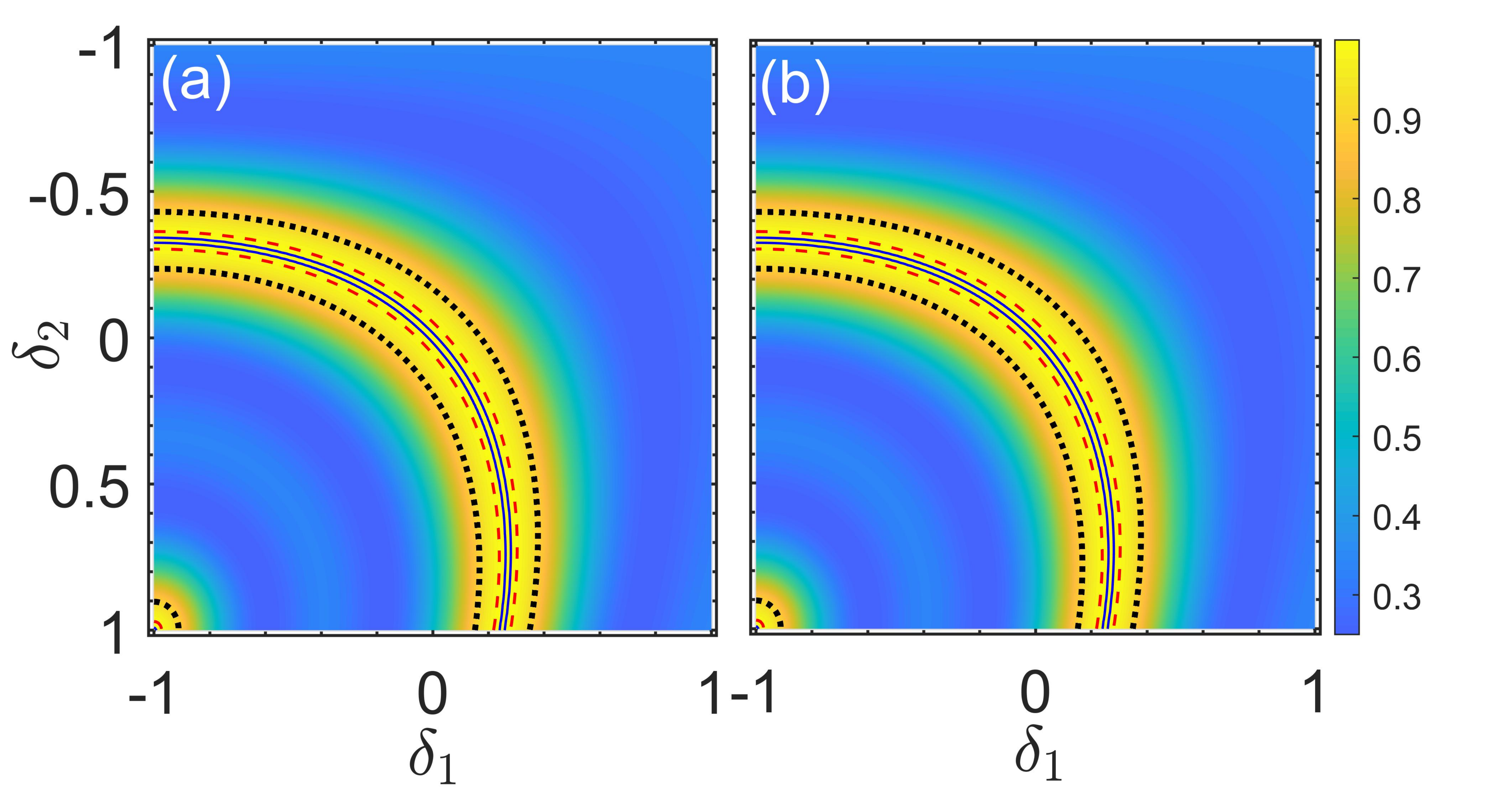}
	\caption{Fidelity $F$ vs the variations $\delta_1$ and $\delta_2$ in the three-pulse sequence, where $\Omega_2=1/\sqrt{3}\Omega_1$, and the phases $\alpha_{12}$ and $\beta_{12}$ are chosen to satisfy (a) Eq.~(\ref{7}) and (b) Eq.~(\ref{8a}), respectively. The results show that the Z gate is less sensitive to the variations in a narrow region. }  \label{fig:02}
\end{figure}

On the other hand, if we set $\Omega_2=1/\sqrt{3}\Omega_1$, we would obtain the Z gate by the three-pulse sequence, which reads
\begin{eqnarray}
\mathbb{U}_{0}=\left[
                \begin{array}{ccc}
                 1 & 0 & 0 \\
                 0 & -1 & 0 \\
                 0  &  0 &  -1
                \end{array}
              \right].
\end{eqnarray}
In Fig.~\ref{fig:02}, we plot the fidelity $F$ of the Z gate as a function of the variations $\delta_1$ and $\delta_2$ in the three-pulse sequence.
Again, the fidelity is robust against the variations in a narrow region.
These results can be explained by the fact that the three-pulse sequence cannot remove the qubit error and the leakage error simultaneously.
In addition, we can observe from Figs.~\ref{fig:01} and \ref{fig:02} that the robustness behaviors of fidelity are quite different when obtaining distinct single-qubit gates.

Next, we consider the case of $N=5$, where the form of the actual evolution operation reads
\begin{eqnarray}
\mathbb{U}_a&=&\exp({-iH_1T})\exp({-iH_2T})\exp({-iH_3T}) \cr
&&\times\exp({-iH_2T})\exp({-iH_1T}). \nonumber
\end{eqnarray}
Since the expressions of the elements $g_j'$ ($j=1,\dots,6$) are too cumbersome to present here, we give the concrete forms in Appendix \ref{b}.
Similarly, in order to eliminate all first-order terms of the error operator $\mathbb{U}_e$, the phases
are demanded to satisfy the following five equations
\begin{equation}   \label{12}
g_1'=\mathrm{Re}(g_2')=\mathrm{Im}(g_2')=\mathrm{Re}(g_{a}')=\mathrm{Im}(g_{a}')=0,
\end{equation}
where $\mathrm{Re}(\cdot)$ and $\mathrm{Im}(\cdot)$ denote the real part and the imaginary part, respectively.

In the five-pulse sequence, there are four variables ($\alpha_{12}$, $\alpha_{23}$, $\beta_{12}$, and $\beta_{23}$) but five equations in Eqs.~(\ref{12}).
Therefore, the phases $\alpha_m$ and $\beta_m$ ($m=1,2,3$) are hardly designed to simultaneously satisfy Eqs.~(\ref{12}) for arbitrary coupling strengths $\Omega_2$ and $\Omega_1$, apart from the case of $\Omega_2=\Omega_1$. Because when $\Omega_2=\Omega_1$, the equation $g_2'=0$ is automatically satisfied. Then, there remain three equations (i.e, $g_1'=g_{a}'=0$) and four variables.
Note that it is difficult for us to achieve the analytical solutions for the phases $\alpha_m$ and $\beta_m$ ($m=1,2,3$), but the numerical solutions are quite readily obtained by solving the remaining three equations. We present one of the solutions in Table~\ref{t1}.
When $\Omega_2\neq\Omega_1$, four variables can satisfy four equations in general. Hence, we divide the solutions into two situations:

($i$) The phases are designed to satisfy the equations $g_1'=g_{a}'=\mathrm{Im}(g_2')=0$.
Remarkably, we hammer at eliminating the leakage error in this situation, and present some of the solutions in Table~\ref{t1}.

($ii$) The phases are designed to satisfy the equations $g_1'=g_{2}'=\mathrm{Im}(g_{a}')=0$.
In this situation, we hammer at eliminating the qubit error, and present some of the solutions in Table~\ref{t2}.

\renewcommand\arraystretch{1.2}
\begin{table}[t]
	\centering
	\caption{Values of the phases in the five-pulse sequence, where the phases are designed to eliminate the leakage error.}
	\label{t1}
	\begin{tabular}{ccc}
		\hline
		\hline
		${\Omega_2}/{\Omega_1}$ & $\alpha_1$~~~$\alpha_2$~~~~~$\alpha_3$~~~ & $\beta_1$~~~$\beta_2$~~~~~$\beta_3$~~~  \\
		\hline
         0.8~~ & ~0~~0.008~~3.574~ & ~0~~0.957~~2.278~ \\
         0.9~~ & ~0~~0.406~~4.275~ & ~0~~1.354~~3.102~ \\
         1.0~~ & ~0~~2.584~~0.000~ & ~0~~4.864~~3.142~ \\
         1.1~~ & ~0~~5.930~~4.121~ & ~0~~0.703~~3.050~ \\
         1.2~~ & ~0~~3.960~~5.856~ & ~0~~2.940~~0.764~ \\
		\hline
		\hline
	\end{tabular}
\end{table}

\renewcommand\arraystretch{1.2}
\begin{table}[b]
	\centering
	\caption{Values of the phases in the five-pulse sequence, where the phases are designed to eliminate the qubit error.}
	\label{t2}
	\begin{tabular}{ccc}
		\hline
		\hline
		 {${\Omega_2}/{\Omega_1}$} & $\alpha_1$~~~$\alpha_2$~~~~~$\alpha_3$~~~ & $\beta_1$~~~$\beta_2$~~~~~$\beta_3$~~~  \\
		\hline
         0.8~~ & ~0~~2.398~~6.033~ & ~0~~4.513~~2.667~ \\
         0.9~~ & ~0~~1.753~~0.432~ & ~0~~3.890~~3.355~ \\
         1.1~~ & ~0~~2.263~~6.077~ & ~0~~4.400~~2.716~ \\
         1.2~~ & ~0~~1.002~~4.533~ & ~0~~5.161~~1.614~ \\
		\hline
		\hline
	\end{tabular}
\end{table}

\begin{figure}[t]
	\centering
	\includegraphics[scale=0.061]{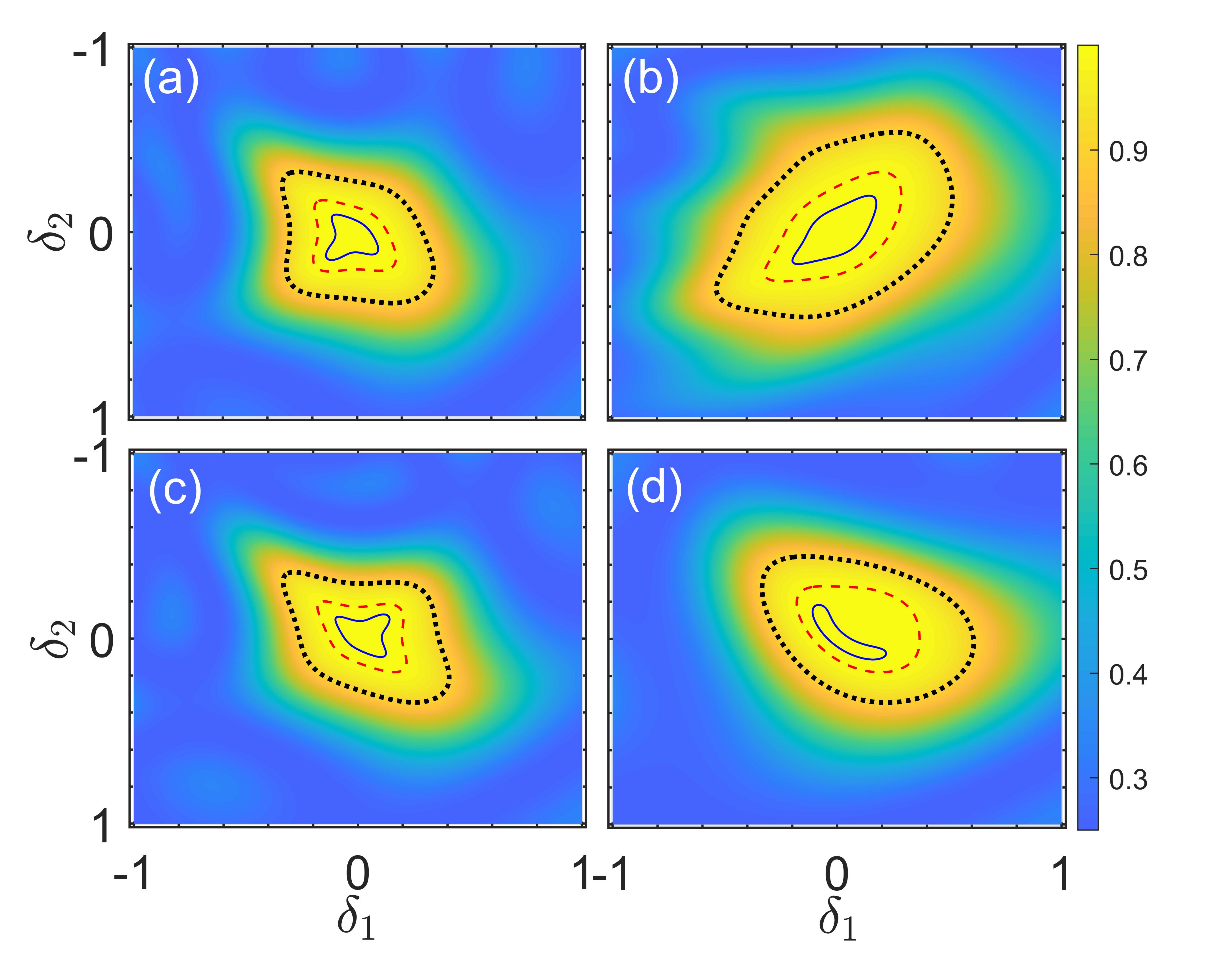}
	\caption{Fidelity $F$ vs the variations $\delta_1$ and $\delta_2$ with different coupling strengths in the five-pulse sequence, where the phases are designed to eliminate the leakage error. (a) ${\Omega_2}/{\Omega_1}=0.9$, (b) ${\Omega_2}/{\Omega_1}=1$, (c) ${\Omega_2}/{\Omega_1}=1.1$, and (d) ${\Omega_2}/{\Omega_1}=1.2$. The fidelity is robust against the variations in a small circular region. }  \label{fig:03}
\end{figure}

\begin{figure}[htbp]
	\centering
 	\includegraphics[scale=0.061]{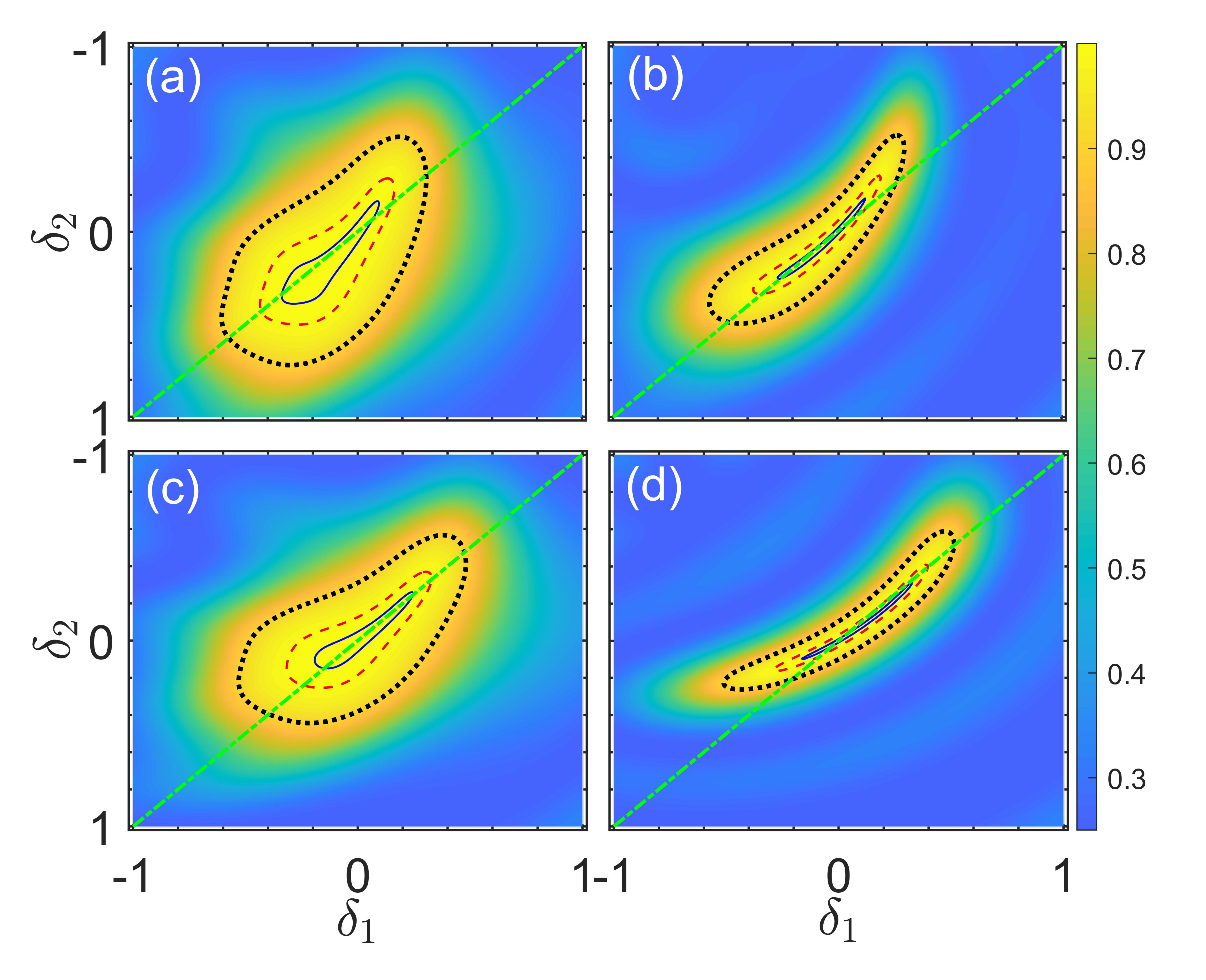}
	\caption{Fidelity $F$ vs the variations $\delta_1$ and $\delta_2$ with different coupling strengths in the five-pulse sequence, where the phases are designed to eliminate the qubit error. (a) ${\Omega_2}/{\Omega_1}=0.8$, (b) ${\Omega_2}/{\Omega_1}=0.9$, (c) ${\Omega_2}/{\Omega_1}=1.1$, and (d) ${\Omega_2}/{\Omega_1}=1.2$. The fidelity is less sensitive to the variations in a narrow strip region. }  \label{fig:04}
\end{figure}

In Fig.~\ref{fig:03}, we plot the fidelity $F$ as a function of the variations $\delta_1$ and $\delta_2$ with different coupling strengths in the five-pulse sequence, where the phases are designed to eliminate the leakage error.
For comparison, the phases in Fig.~\ref{fig:04} are designed to eliminate the qubit error.
First of all, we observe in Fig.~\ref{fig:03}(b) that the fidelity is significantly improved and robust against the variations in a wide region when $\Omega_2=\Omega_1$, since all first-order terms of the error operator $\mathbb{U}_e$ are removed at the same time in the five-pulse sequence.
Second, the robustness behaviors of fidelity are quite different in the two situations.
When we dedicate to eliminating the leakage error, the fidelity is robust against the variations in a small circular region.
While we dedicate to eliminating the qubit error, the fidelity is robust against the variations in a very narrow strip region around $\delta_1=-\delta_2$ [the green (light gray) line in Fig.~\ref{fig:04}].

Finally, we consider the case of $N=7$, where the actual evolution operator can be written as
\begin{eqnarray}
\mathbb{U}_a&=&\exp({-iH_1T})\exp({-iH_2T})\exp({-iH_3T})\exp({-iH_4T}) \cr
&&\times\exp({-iH_3T})\exp({-iH_2T})\exp({-iH_1T}). \nonumber
\end{eqnarray}
In this case, there are six variables ($\alpha_{12}$, $\alpha_{23}$, $\alpha_{34}$, $\beta_{12}$, $\beta_{23}$, and $\beta_{34}$) that can be used to eliminate errors.
For the first-order terms of the error operator $\mathbb{U}_e$, there are only five constraint equations. Thus, it is possible to eliminate the first-order error by the seven-pulse sequence. What is more, when $\Omega_2=\Omega_1$, three constraint equations remain. Hence, we can further eliminate the higher-order term of the error operator $\mathbb{U}_e$, e.g., the term $\delta_1\delta_2$.

\begin{figure}[htbp]
	\centering
	\includegraphics[scale=0.07]{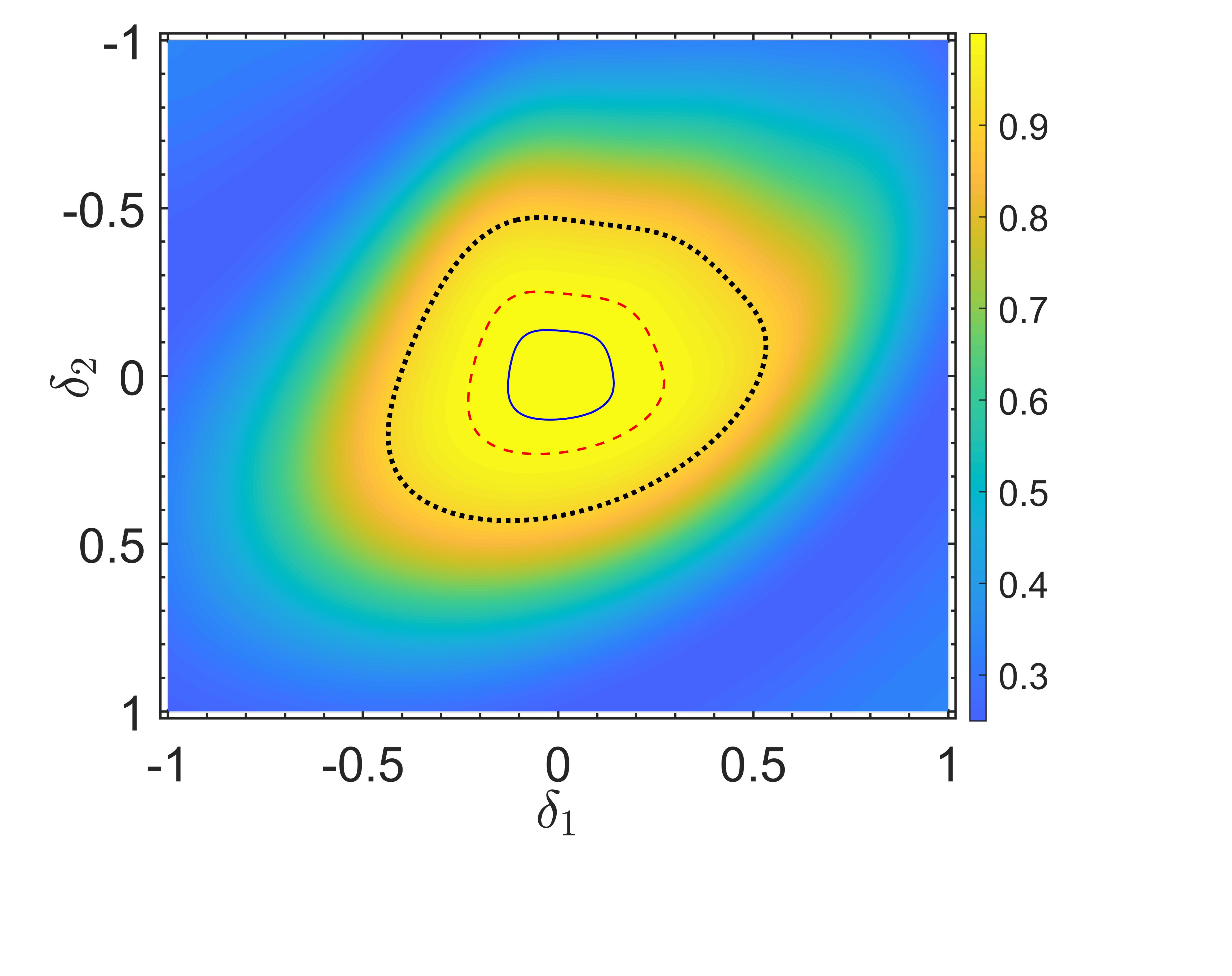}
	\caption{Fidelity $F$ vs the variations $\delta_1$ and $\delta_2$ in the seven-pulse sequence, where ${\Omega_2}/{\Omega_1}=1$ and $\varphi_7=0.948$. It shows that the robustness region is wider than that of the five-pulse sequence. }  \label{fig:05}
\end{figure}

Here, we consider $\Omega_1=\Omega_2$ in the seven-pulse sequence for simplicity. Because of its complexity, the expression of the actual evolution operator is given in Appendix \ref{c}. Again, we would obtain the NOT gate, and it has the following form
\begin{eqnarray}
\mathbb{U}_{0}=\left[
                \begin{array}{ccc}
                 0 & \exp({-i\varphi_7}) & 0 \\
                 \exp({i\varphi_7}) & 0 & 0 \\
                 0 & 0 & \exp({i\Upsilon_7})
                \end{array}
              \right],
\end{eqnarray}
where $\varphi_7=-2(\gamma_1-\gamma_2+\gamma_3)+\gamma_4$.
The phases $\alpha_m$ and $\beta_m$ ($m=1,2,3,4$) are designed to satisfy the following equations
\begin{equation}
g'_1=\mathrm{Re}(g'_a)=\mathrm{Im}(g'_a)=\mathrm{Re}(g'_9)=\mathrm{Im}(g'_9)=g'_7=0.
\end{equation}
We present some numerical solutions in Table~\ref{t3}, and plot
the fidelity $F$ as a function of the variations $\delta_1$ and $\delta_2$ in Fig.~\ref{fig:05} when $\varphi_7=0.948$. An inspection of Fig.~\ref{fig:05} demonstrates that the fidelity has a wider robustness region when comparing to the case of the five-pulse sequence [see Fig.~\ref{fig:03}(b)]. Particularly, it is robust against the variations along all directions.
Note that with the increasing of the number of pulses, the higher-order terms in the error operator $\mathbb{U}_e$ can be effectively removed. As a result, the fidelity would have a wider and wider robustness region.

\renewcommand\arraystretch{1.2}
\begin{table}[htbp]
	\centering
	\caption{Values of the phases in the seven-pulse sequence, where $\Omega_2/\Omega_1=1$.}
	\label{t3}
	\begin{tabular}{ccc}
		\hline
		\hline
	$\alpha_1$~~~~$\alpha_2$~~~~~$\alpha_3$~~~~~$\alpha_4$~~~ & $\beta_1$~~~$\beta_2$~~~~~$\beta_3$~~~~~~$\beta_4$~~~ & $\varphi_7$ \\
		\hline
         0.0~~4.117~~2.286~~5.641~ & ~0~~3.165~~0.920~~3.864~ &~0.948 \\
         0.3~~4.406~~2.438~~5.685~ & ~0~~3.190~~0.870~~3.723~ &~0.657 \\
         0.5~~3.642~~1.536~~4.662~ & ~0~~4.181~~2.608~~6.168~ &~--1.442 \\
         0.8~~4.102~~1.580~~4.160~ & ~0~~4.158~~1.846~~4.770~ &~--1.789 \\
         1.0~~3.915~~2.385~~5.904~ & ~0~~3.705~~4.822~~1.086~ &~1.829 \\
		\hline
		\hline
	\end{tabular}
\end{table}

%

\section{Applications}  \label{IV}

In this section, we employ the above composite-pulses theory to implement robust quantum information transfer between two qubits \cite{PhysRevLett.91.207901,PhysRevLett.92.187902,PhysRevA.81.012304,PhysRevA.91.032301,PhysRevA.102.012418,Yousefjani20}. To be specific, suppose that the state of the first qubit is $|\Phi\rangle_1=\cos\Theta|1\rangle_{1}+\exp({i\vartheta})\sin\Theta|2\rangle_{1}$, where $\Theta$ and $\vartheta$ can be arbitrary values.
The goal is to transfer $|\Phi\rangle_1$ to the second qubit, that is, to let the state of the second qubit become $|\Phi\rangle_2=\cos\Theta|1\rangle_{2}+\exp({i\vartheta})\sin\Theta|2\rangle_{2}$.

Consider the atom-cavity system in which two identical atoms with $\Lambda$-type three-level structure are trapped in an optical cavity. The atom has two ground states $|1\rangle_{k}$ and $|2\rangle_{k}$, and an excited state $|3\rangle_{k}$ (the subscript $k$ denotes the $k$th atom), where the two ground states acted as a qubit.
The transitions $|1\rangle_{k}\leftrightarrow|3\rangle_{k}$ are resonantly driven by the laser fields with the coupling strength $\Omega_k$ and the phase $\alpha_k$. The transitions $|2\rangle_{k}\leftrightarrow|3\rangle_{k}$ are resonantly coupled to the cavity mode with the coupling constant $g_k$.
Therefore, in the interaction picture, the Hamiltonian of the atom-cavity system reads ($\hbar=1$)
\begin{equation}
H=\sum_{k=1}^{2}\Big[(1+\delta_k)\Omega_k\exp({i\alpha_k})|1\rangle_{kk}\langle 3|+g_k|3\rangle_{kk}\langle 2|a+\mathrm{H.c.}\Big],
\end{equation}
where $a$ is the annihilation operator of the cavity mode, and $\delta_k$ are the variations due to the spatial inhomogeneity of the laser fields. Note that this physical model can also be found in the diamond nitrogen-vacancy centers coupled to the whispering-gallery mode of a microsphere cavity \cite{Yang2010} or the trapped ion-cavity system \cite{PhysRevA.85.062308}. For simplicity, we suppose that $\Omega_k=\Omega$, $\delta_k=\delta$, and $g_k=g$.

In this system, quantum information transfer between two qubits can be illustrated as
\begin{eqnarray} \label{201}
&&|\Psi_i\rangle=(\cos\Theta|1\rangle_{1}+\exp({i\vartheta})\sin\Theta|2\rangle_{1})|2\rangle_2|0\rangle_c \cr
&&\xLongrightarrow[\mathrm{pulses}]{\textrm{composite}} |\Psi_f\rangle=|2\rangle_1(\cos\Theta|1\rangle_{2}+\exp({i\vartheta})\sin\Theta|2\rangle_{2})|0\rangle_c, \nonumber\\
\end{eqnarray}
where $|\Psi_i\rangle$ is the initial state of the system, $|\Psi_f\rangle$ is the final state after the composite pulses, and $|m\rangle_c$ represents that there are $m$ photons in the cavity. Once the equation $|\Psi_f\rangle=\mathbb{U}_{0}|\Psi_i\rangle$ is satisfied, it means that we successfully transfer the quantum information from the first qubit to the second qubit. Here, $\mathbb{U}_{0}$ is the evolution operator of the atom-cavity system by CPs.

It is \textcolor[rgb]{0.00,0.00,1.00}{easy} to verify that the excited number operator $N_e$ of this system is a conserved quantity, where $N_e=\sum_k(|1\rangle_{kk}\langle 1|+|3\rangle_{kk}\langle 3|)+a^{\dag}a$. Hence, the system state remains unchanged by CPs if the initial state is $|2\rangle_1|2\rangle_2|0\rangle_c$, because the null excited subspace (i.e., $N_e=0$) only contains the state $|2\rangle_1|2\rangle_2|0\rangle_c$. In short, it can be expressed by
\begin{eqnarray} \label{220}
|2\rangle_1|2\rangle_2|0\rangle_c=\mathbb{U}_{0}|2\rangle_1|2\rangle_2|0\rangle_c.
\end{eqnarray}
Therefore, if we obtain the population inversion from $|1\rangle_1|2\rangle_2|0\rangle_c$ to $|2\rangle_1|1\rangle_2|0\rangle_c$ by CPs, i.e.,
\begin{eqnarray}
|2\rangle_1|1\rangle_2|0\rangle_c=\mathbb{U}_{0}|1\rangle_1|2\rangle_2|0\rangle_c,
\end{eqnarray}
according to expression~(\ref{201}), the quantum information transfer between two qubits is naturally implemented.
Note that the relative phase $\vartheta$ can be handled by properly choosing the values of $\alpha_1$ and $\beta_1$.
In the following, we will focus on the population inversion from $|1\rangle_1|2\rangle_2|0\rangle_c$ to $|2\rangle_1|1\rangle_2|0\rangle_c$, since Eq.~(\ref{220}) always sets up by CPs.

Due to the conservation of the excited number, we can restrict the system dynamics into the single excited subspace, i.e., $N_e=1$. When $g\gg\Omega$, the effective Hamiltonian of the atom-cavity system can be written as \cite{PhysRevA.96.022314}
\begin{eqnarray} \label{211}
H_{\mathrm{eff}}&=&\Big[\frac{(1+\delta)\Omega \exp({i\alpha_1})}{\sqrt{2}}|\Phi_1\rangle\langle \Phi_2| \cr
&&-\frac{(1+\delta)\Omega \exp({i\alpha_2})}{\sqrt{2}} |\Phi_1\rangle\langle \Phi_3|\Big]+\mathrm{H.c.},
\end{eqnarray}
where $|\Phi_1\rangle={1}/{\sqrt{2}}\big(|2\rangle_1|3\rangle_2|0\rangle_c-|3\rangle_1|2\rangle_2|0\rangle_c\big)$,
$|\Phi_2\rangle=|2\rangle_1|1\rangle_2|0\rangle_c$, and $|\Phi_3\rangle=|1\rangle_1|2\rangle_2|0\rangle_c$.
We find from Eq.~(\ref{211}) that the effective Hamiltonian in the single excited subspace is equivalent to the Hamiltonian of the three-level physical model. Therefore, by designing the phases of the laser fields to achieve the NOT gate as studied in Sec.~\ref{III}, we can implement population inversion from $|1\rangle_1|2\rangle_2|0\rangle_c$ to $|2\rangle_1|1\rangle_2|0\rangle_c$ in a robust manner.

\begin{figure}[t]
	\centering
	\includegraphics[scale=0.432]{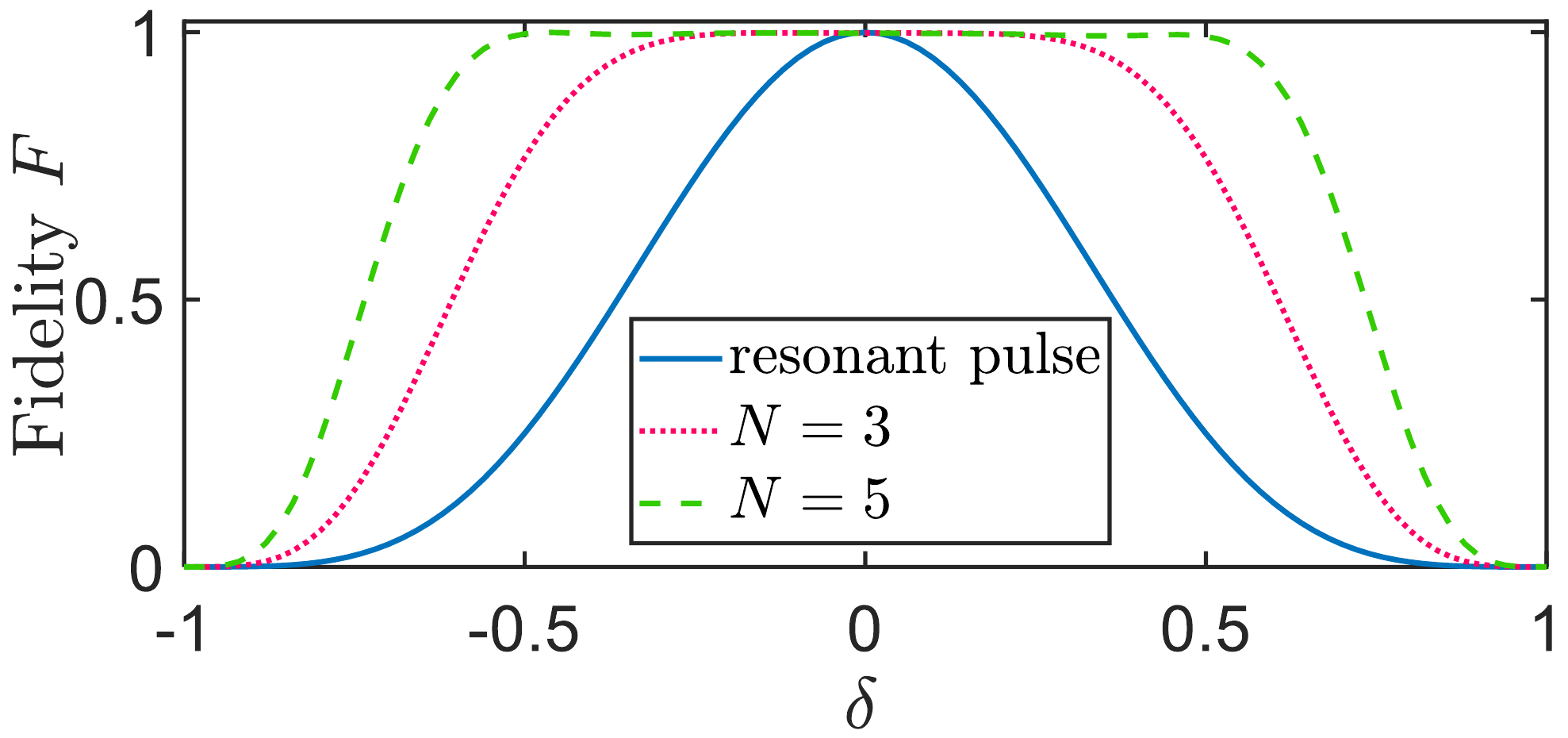}
	\caption{Fidelity $F$ vs the variation $\delta$, where the fidelity is defined by $F=|\langle\Psi_t|1\rangle_1|2\rangle_2|0\rangle_c|^2$ and $|\Psi_t\rangle$ is the system state after the composite pulses. $g=30\Omega$. The solid blue curve, the dotted red curve, and the dashed green curve correspond to the resonant pulse, the three-pulse sequence, and the five-pulse sequence, respectively. It shows that the quantum information transfer is robust in the five-pulse sequence, because $F>0.992$ even when $|\delta|=0.5$.}  \label{fig:06}
\end{figure}

In Fig.~\ref{fig:06}, we plot the fidelity $F$ as a function of the variation $\delta$ by CPs, where the fidelity is defined by $F=|\langle\Psi_t|1\rangle_1|2\rangle_2|0\rangle_c|^2$ and $|\Psi_t\rangle$ is the system state after CPs. For comparison, we also plot the fidelity of the final state by the resonant pulse.
Remarkably, the fidelity is robust against the variations of the laser fields by the composite pulses, especially in the five-pulse sequence. In this case, the fidelity still maintains a high value ($>$$99.2\%$) even though the variations of the coupling strengths reach $\pm50\%$.

\begin{figure}[htbp]
	\centering
	\includegraphics[scale=0.053]{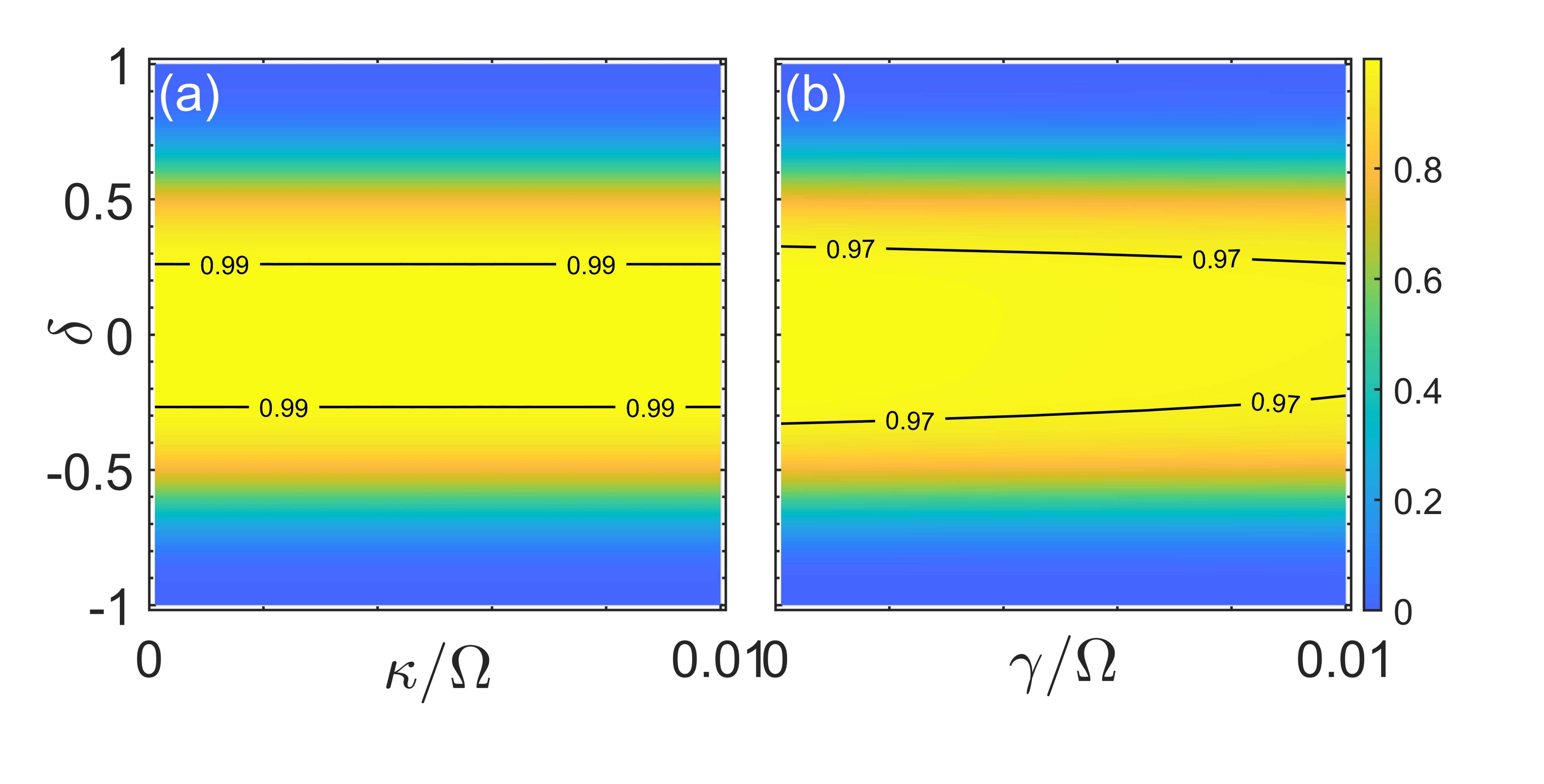}
	\caption{(a) Fidelity $F$ vs the variation $\delta$ and the decay rate $\kappa$ in the three-pulse sequence, where $\gamma=0$. (b) Fidelity $F$ vs the variation $\delta$ and the dissipation rate $\gamma$ in the three-pulse sequence, where $\kappa=0$. The other parameters are the same as in Fig.~\ref{fig:06}. One can observe that $F \gtrsim 0.99$ for $\kappa\lesssim0.01$ and $\delta\lesssim0.25$ in (a), while $F \gtrsim 0.97$ for $\gamma\lesssim0.01$ and $\delta\lesssim0.25$ in (b). The results demonstrate that the decoherence of the excited states is dominant during the operation process.}  \label{fig:08}
\end{figure}

So far, we do not consider the influence of decoherence on the fidelity of population inversion. In the presence of decoherence induced by the atomic spontaneous emission and the cavity decay, the master equation of the whole system can be written in the Lindblad form,
\begin{eqnarray}
\dot{\rho}&=&-i[H,\rho]+\sum_{k,l=1}^{2}\frac{\gamma_{kl}}{2}(\sigma_{l3}^{k}\rho\sigma_{3l}^{k}- \sigma_{3l}^{k}\sigma_{l3}^{k}\rho-\rho\sigma_{3l}^{k}\sigma_{l3}^{k}) \cr
&&+\frac{\kappa}{2}(2a\rho a^{\dag}-a^{\dag}a\rho-\rho a^{\dag}a),
\end{eqnarray}
where $\sigma_{l3}^{k}=|l\rangle_{kk}\langle3|$, $\gamma_{kl}$ is the dissipation rate of the $k$th atom from the excited state $|3\rangle$ to the ground state $|l\rangle$, and $\kappa$ is the decay rate of the cavity mode. For simplicity, we assume $\gamma_{k1}=\gamma_{k2}=\gamma/2$.
In Fig.~\ref{fig:08}(a), we observe that the protocol is robust against the cavity decay, since the fidelity keeps almost a high value ($F \gtrsim 0.99$) when the variation is not very large.
On the other hand, the dissipation of the excited state is the main decoherence of this system, as shown in Fig.~\ref{fig:08}(b). This is because the transition between the ground states and the excited state is resonantly driven by the laser pulses.
An alternative method of decreasing the decoherence of the excited state is to adopt the Raman process (the so-called two-photon resonance); i.e., both ground states are coupled to the excited state with large detuning. As a result, one can adiabatically eliminate the excited state. Nevertheless, this detuned interaction would prolong the manipulation time, and we should make a trade-off between the detuning and the manipulation time.

Finally, we briefly demonstrate the specific physical systems that can be implemented in experiments. First of all, the physical model can be found in either the cesium atoms trapped in the optical cavity or the diamond nitrogen-vacancy centers coupled to the whispering-gallery mode of a microsphere cavity \cite{Yang2010}.
In the cesium atom, the hyperfine states $|F=3,m=2\rangle$ and $|F=4,m=4\rangle$ of the $6^2S_{1/2}$ electronic state corresponds to the ground states $|1\rangle$ and $|2\rangle$, respectively, and the hyperfine state $|F=3,m=3\rangle$ of the $6^2P_{1/2}$ serves as the excited state $|3\rangle$.
In the diamond nitrogen-vacancy centers \cite{Yang2010}, the states $|^3A,m_s=0\rangle$ and $|^3A,m_s=-1\rangle$ act as the ground states $|1\rangle$ and $|2\rangle$, respectively, and the state $|^3E,m_s=0\rangle$ corresponds to the excited state $|3\rangle$.
Note that the coupling strength between the nitrogen-vacancy centers and the laser fields can reach $2\pi\times100$MHz \cite{Yang2010}.
For the composite pulse sequence, the waveform is actually the square-wave pulse. This is easily produced by a pulse shaper in experiments. In order to modulate the phases of laser fields, one can employ a phase modulator to generate distinct phases \cite{Wollenhaupt2005,cho2008,chen2010,keating2016,zhang17,shi17}. Then, the total manipulation time is about 17.68 ns in the five-pulse sequence.

\section{Conclusion} \label{V}

In summary, we have presented a general method to implement robust universal single-qubit gates by composite pulses in $\Lambda$-type three-level systems. The basic idea is to expand the actual evolution operator by Taylor expansion,
then remove the first (higher)-order terms of the error operator by designing distinct phases of the coupling strengths. As a result, we would obtain a robust single-qubit gate.

In the three-pulse sequence, we can eliminate either the qubit error or the leakage error, but not both errors, because there are only two variables to control. Thus, the fidelity is less sensitive to the variations only along a specific direction. In the five-pulse sequence, since there are four variables, we can eliminate all first-order terms of the error operator when the coupling strengths are the same, and the fidelity is robust against the variations in a wide region. Additionally, the robustness behaviors are quite different for distinct phases. In the seven-pulse sequence, there are more variables to control; thus we can eliminate the higher-order terms of the error operator. As a result, it is robust against the variations in a wider region than that of the five-pulse sequence.
What is more, we have exemplified the composite pulses method to implement robust quantum information transfer between two qubits.
Finally, the emphasis of this work is on the compensation of deviations in three-level systems, but it is possible to generalize in the quantum systems with distinct structure, such as the four-level configuration \cite{PhysRevA.97.013830}, the Raman-type configuration \cite{PhysRevResearch.2.043194}, etc. Those results should be helpful to robust control in complicated quantum systems.
\\

\begin{acknowledgments}
	This work is supported by the National Natural Science Foundation of China under Grant No. 11805036, No. 11674060, No. 11534002, and No. 11775048.
\end{acknowledgments}

\begin{widetext}

\begin{appendix}

\section{The elements for the operator error in case of $N=5$}  \label{b}

In this Appendix, we present the expressions for the elements $g_j'$ ($j=1,\dots,6$) when we choose $N=5$. Namely,
\begin{eqnarray}
g_1'&=&\frac{4\Omega_1^2\Omega_2^2}{(\Omega_1^2+\Omega_2^2)^6}\Big[a_0+a_1\cos(\alpha_{12}-\beta_{12}) +a_2\cos(\alpha_{12}-\alpha_{23}-\beta_{12}+\beta_{23}) +a_3\cos(\alpha_{13}-\beta_{13})+a_4\cos(\alpha_{23}-\beta_{23})\Big],  \cr\cr
g_2'&=&\frac{2\exp({i\gamma_1})\Omega_1\Omega_2(\Omega_1^2-\Omega_2^2)}{(\Omega_1^2+\Omega_2^2)^6} \bigg\{80\Omega_1^4\Omega_2^4\exp\big[{i(\alpha_{12}-\alpha_{23}-\beta_{12}+\beta_{23})}\big] +b_1\big[\exp({i\beta_{13}}-i\alpha_{13})- 2\exp({i\beta_{12}-i\alpha_{12}})\big] \cr&&+b_2\Big[4\cos{(\alpha_{23}-\beta_{23})}+2\exp({i\alpha_{12}-i\beta_{12}}) -\exp[{-i(\alpha_{12}-\alpha_{23}-\beta_{12}+\beta_{23})}]-\exp({i\alpha_{13}-i\beta_{13}})\Big] +2b_3\bigg\},  \cr\cr
g_3'&=&\frac{\Omega_1^2}{\Omega_2^2}g_5'=\frac{i\exp({i\alpha_1})\pi\Omega_1^3}{(\Omega_1^2+\Omega_2^2)^{11/2}} g_{a}' g_{b}',  \cr\cr
g_4'&=&\frac{\Omega_2^2}{\Omega_1^2}g_6'=\frac{i\exp({i\beta_1})\pi\Omega_1^2\Omega_2}{(\Omega_1^2+\Omega_2^2)^{11/2}} g_{a}' g_{c}',  \nonumber
\end{eqnarray}
where $\alpha_{mn}=\alpha_m-\alpha_n$,  $\beta_{mn}=\beta_m-\beta_n$, $m,n=1,2,3$, and
\begin{eqnarray}
a_0&=&-9\Omega_1^8+76\Omega_1^6\Omega_2^2-150\Omega_1^4\Omega_2^4+76\Omega_1^2\Omega_2^6-9\Omega_2^8,   \cr\cr
a_1&=&8(\Omega_1^2-\Omega_2^2)^2(\Omega_1^4-8\Omega_1^2\Omega_2^2+\Omega_2^4),   \cr\cr
a_2&=&32\Omega_1^2\Omega_2^2(\Omega_1^4-3\Omega_1^2\Omega_2^2+\Omega_2^4),   \cr\cr
a_3&=&4(\Omega_1^2-\Omega_2^2)^2(\Omega_1^4-8\Omega_1^2\Omega_2^2+\Omega_2^4),   \cr\cr
a_4&=&4(\Omega_1^8-18\Omega_1^6\Omega_2^2+42\Omega_1^4\Omega_2^4-18\Omega_1^2\Omega_2^6+\Omega_2^8),   \cr\cr
b_1&=&(\Omega_1^2-\Omega_2^2)^2(\Omega_1^4-18\Omega_1^2\Omega_2^2+\Omega_2^4),   \cr\cr
b_2&=&4\Omega_1^2\Omega_2^2(3\Omega_1^4-14\Omega_1^2\Omega_2^2+3\Omega_2^4),   \cr\cr
b_3&=&\Omega_1^8-32\Omega_1^6\Omega_2^2+94\Omega_1^4\Omega_2^4-32\Omega_1^2\Omega_2^6+\Omega_2^8,   \cr\cr
g_{a}'&=&4\Omega_1^2\Omega_2^2\big[\exp({i\alpha_{12}+i\beta_{23}})+ \exp({i\alpha_{23}+i\beta_{12}})\big]+2(\Omega_1^2-\Omega_2^2)\big[\Omega_1^2\exp({i\alpha_{13}})- \Omega_2^2\exp({i\beta_{13}})\big] \cr &&+2(\Omega_1^2+\Omega_2^2)\big[\Omega_1^2\exp({i\alpha_{23}})+\Omega_2^2\exp({i\beta_{23}})\big]+(\Omega_1^2+\Omega_2^2)^2, \cr\cr
g_{b}'&=& 2\Omega_2^2(\Omega_1^2-\Omega_2^2)\big[\exp({-i\alpha_{12}-i\beta_{23}})- \exp({-i\beta_{13}})\big]+4\Omega_1^2\Omega_2^2\exp({-i\alpha_{23}-i\beta_{12}})+(\Omega_1^2-\Omega_2^2)^2\exp({-i\alpha_{13}}), \cr\cr
g_{c}'&=& 2\Omega_1^2(\Omega_1^2-\Omega_2^2)\big[\exp({-i\alpha_{13}})-\exp({-i\alpha_{23}-i\beta_{12}}) \big]+4\Omega_1^2\Omega_2^2\exp({-i\alpha_{12}-i\beta_{23}})+(\Omega_1^2-\Omega_2^2)^2\exp({-i\beta_{13}}),  \nonumber
\end{eqnarray}

\section{The expressions for the actual evolution operator in case of $N=7$}  \label{c}

In this Appendix, we present the expression for the actual evolution operator $\mathbb{U}_a$ when we choose $N=7$ and $\Omega_1=\Omega_2$, namely,
\begin{eqnarray}  \label{b4}
\mathbb{U}_a&=&\mathbb{U}_0+\mathbb{U}_e \cr\cr
            &=&\left[
                \begin{array}{ccc}
                 0 & e^{-i\varphi_7} & 0 \\
                 e^{i\varphi_7} & 0 & 0 \\
                 0 & 0 & e^{i\Upsilon_7}
                \end{array}
              \right]+\left[
                \begin{array}{ccc}
                 g'_1 & -g'_2 & g'_3 \\
                 -g^{\prime *}_2 & -g'_1 & g'_4 \\
                 -g^{\prime *}_3 & -g^{\prime *}_4 & 0
                \end{array}
              \right]\delta_1
+\left[
                \begin{array}{ccc}
                 -g'_1 & g'_2 & g'_5 \\
                 g^{\prime *}_2 & g'_1 & g'_6 \\
                 -g^{\prime *}_5 & -g^{\prime *}_6 & 0
                \end{array}
              \right]\delta_2+\left[
                \begin{array}{ccc}
                 g'_7 & g'_9 & g'_{10} \\
                 g^{\prime *}_{9} & g'_8 & g'_{11} \\
                 -g^{\prime *}_{10} & -g^{\prime *}_{11} & 0
                \end{array}
              \right]\delta_1\delta_2+\mathcal{O}(\delta_1^2,\delta_2^2),  \nonumber
\end{eqnarray}
where
\begin{eqnarray}
\varphi_7&=&-2(\gamma_1-\gamma_2+\gamma_3)+\gamma_4,   \cr
g'_1&=&-1+2\cos(\alpha_{34}-\beta_{34})+2\cos(\alpha_{12}-\alpha_{23}+\alpha_{34} -\beta_{12}+\beta_{23}-\beta_{34})-2\cos(\alpha_{23}-\alpha_{34}-\beta_{23}+\beta_{34}), \cr
g'_2&=&0, \cr
g'_3&=&g'_5=2g'_{10}=-\frac{i\pi \exp\big[{i(\alpha_{4}+\alpha_{12}+\alpha_{34}-\beta_{12}-\beta_{34})}\big]}{2\sqrt{2}}g'_a, \cr
g'_4&=&g'_6=2g'_{11}=-\frac{i\pi \exp\big[{i(\beta_{4}-\alpha_{12}-\alpha_{34}+\beta_{12}+\beta_{34})}\big]}{2\sqrt{2}}g'_a, \cr
g'_7&=&-\frac{\pi^2}{8}\Big[7+4\cos\alpha_{23}+4\cos\beta_{12}+4\cos\beta_{34}+ 4\cos(\alpha_{23}+\beta_{34})+ 4\cos(\alpha_{23}+\beta_{12})+2\cos(\alpha_{34}-\beta_{34})  \cr
&&+4\cos(\alpha_{23}-\alpha_{34}+\beta_{34})+4\cos(\alpha_{23}+\beta_{12}+\beta_{34}) +4\cos(\alpha_{23}-\alpha_{34}+\beta_{12}-\beta_{23}+\beta_{34})  \cr
&&+2\cos(\alpha_{23}-\alpha_{34}-\beta_{23}+\beta_{34}) +4\cos(\alpha_{23}-\alpha_{34}+\beta_{12}+\beta_{34})
+2\cos(\alpha_{12}-\alpha_{23}+\alpha_{34}-\beta_{12}+\beta_{23}-\beta_{34})  \Big], \cr
g'_8&=&-\frac{\pi^2}{8}\Big[7+4\cos\beta_{23}+4\cos\alpha_{12}+4\cos\alpha_{34}+ 4\cos(\beta_{23}+\alpha_{34})+ 4\cos(\beta_{23}+\alpha_{12})+2\cos(\alpha_{34}-\beta_{34})  \cr
&&+4\cos(\beta_{23}-\beta_{34}+\alpha_{34})+4\cos(\beta_{23}+\alpha_{12}+\alpha_{34}) +2\cos(\alpha_{12}-\alpha_{23}+\alpha_{34}-\beta_{12}+\beta_{23}-\beta_{34})  \cr
&&+2\cos(\beta_{23}-\beta_{34}-\alpha_{23}+\alpha_{34})+4\cos(\beta_{23}-\beta_{34}+\alpha_{12}-\alpha_{23}+\alpha_{34}) +4\cos(\beta_{23}-\beta_{34}+\alpha_{12}+\alpha_{34})     \Big],   \cr
g'_9&=&\frac{1}{8}\exp\big[{i(\gamma_4-2\beta_{14})}\big]\Big\{ 32\exp\big[i(\alpha_{14}+\alpha_{34}+\beta_{13})\big]- 32\exp\big[i(2\alpha_{14}-\alpha_{23}+\beta_{23})\big] \cr&&-32\exp\big[i(\alpha_{12}+2\alpha_{34}+\beta_{13}+\beta_{23})\big] +16\exp\big[i2(\alpha_{13}+\beta_{34})\big] +16\exp\big[i2(\alpha_{23}+\beta_{12}+\beta_{34})\big] \cr&&
+16\exp\big[i2(\alpha_{12}+\beta_{24})\big] -32\exp\big[i(\alpha_{13}+\alpha_{23}+\beta_{12}+2\beta_{34})\big] -32\exp\big[i(\alpha_{12}+\alpha_{13}+\beta_{24}+\beta_{34})\big] \cr&& +32\exp\big[i(\alpha_{13}+\beta_{14}+\beta_{34})\big] -2\pi^2\exp\big[i(\alpha_{12}+\alpha_{34}+\beta_{23})\big] -2\pi^2\exp\big[i(2\alpha_{12}+\alpha_{34}+\beta_{23})\big] \cr&& -2\pi^2\exp\big[i(\alpha_{12}+2\alpha_{34}+\beta_{23})\big] -2\pi^2\exp\big[i(2\alpha_{12}+2\alpha_{34}+\beta_{23})\big] -2\pi^2\exp\big[i(2\alpha_{12}+\alpha_{34}+2\beta_{34})\big] \cr&& -2\pi^2\exp\big[i(2\alpha_{12}+2\alpha_{34}+\beta_{13}+\beta_{23})\big] -2\pi^2\exp\big[i(2\alpha_{14}-\alpha_{23}+\beta_{13}+\beta_{23})\big] \cr&&-2\pi^2\exp\big[i(\alpha_{12}+2\alpha_{34}+2\beta_{23})\big]
-2\pi^2\exp\big[i(2\alpha_{14}-\alpha_{23}+2\beta_{23})\big]
-2\pi^2\exp\big[i(\alpha_{12}+\alpha_{34}+\beta_{24})\big] \cr&&
-2\pi^2\exp\big[i(2\alpha_{12}+\alpha_{34}+\beta_{24})\big] -2\pi^2\exp\big[i(\alpha_{14}+\beta_{24})\big] -2\pi^2\exp\big[i(\alpha_{12}+\alpha_{14}+\beta_{14})\big] \cr&& -2\pi^2\exp\big[i(\alpha_{12}+\alpha_{14}+\beta_{23}+\beta_{24})\big] -2\pi^2\exp\big[i(2\alpha_{12}+2\alpha_{34}+\beta_{23}+\beta_{24})\big] \cr&&-2\pi^2\exp\big[i(2\alpha_{14}-\alpha_{23}+\beta_{23}+\beta_{24})\big] -2\pi^2\exp\big[i(\alpha_{12}+\alpha_{14}+\beta_{14}+\beta_{23})\big] \cr&&-2\pi^2\exp\big[i(2\alpha_{14}-\alpha_{23}+\beta_{14}+\beta_{23})\big] -2(\pi^2-16)\exp\big[i(\alpha_{12}+\alpha_{14}+\beta_{24})\big] -7(\pi^2-8)\exp\big[i2(\alpha_{12}+\alpha_{34}+\beta_{23})\big] \cr&&-2(\pi^2+16)\exp\big[i(\alpha_{14}+\beta_{14})\big] -2(\pi^2+16)\exp\big[i(2\alpha_{12}+\alpha_{34}+\beta_{23}+\beta_{24})\big] \Big\}, \cr
g'_a&=&1\!+\! \exp({i\alpha_{34}})\!+\!\exp({i\beta_{34}})\!+\!\exp[{i(\alpha_{34}+\beta_{23})}]\!+\!\exp[{i(\beta_{34}+\alpha_{23})}] \!+\!\exp[{i(\alpha_{12}+\alpha_{34}+\beta_{23})}]\!+\!\exp[{i(\beta_{12}+\beta_{34}+\alpha_{23})}].  \nonumber
\end{eqnarray}
Here, $\alpha_{mn}=\alpha_{m}-\alpha_{n}$ and $\beta_{mn}=\beta_{m}-\beta_{n}$.
In order to eliminate the first-order and higher-order terms of the error operator $\mathbb{U}_e$, one should satisfy the following equations:
\begin{eqnarray}
g'_1=\mathrm{Re}(g'_a)=\mathrm{Im}(g'_a)=\mathrm{Re}(g'_9)=\mathrm{Im}(g'_9)=g'_7=g'_8=0.  \nonumber
\end{eqnarray}
There are seven equations but only six variables. Hence, we ignore the last equation for the simulations.

\end{appendix}

\end{widetext}

\bibliographystyle{apsrev4-1}
\bibliography{references}

\begin{thebibliography}{92}%
\makeatletter
\providecommand \@ifxundefined [1]{%
 \@ifx{#1\undefined}
}%
\providecommand \@ifnum [1]{%
 \ifnum #1\expandafter \@firstoftwo
 \else \expandafter \@secondoftwo
 \fi
}%
\providecommand \@ifx [1]{%
 \ifx #1\expandafter \@firstoftwo
 \else \expandafter \@secondoftwo
 \fi
}%
\providecommand \natexlab [1]{#1}%
\providecommand \enquote  [1]{``#1''}%
\providecommand \bibnamefont  [1]{#1}%
\providecommand \bibfnamefont [1]{#1}%
\providecommand \citenamefont [1]{#1}%
\providecommand \href@noop [0]{\@secondoftwo}%
\providecommand \href [0]{\begingroup \@sanitize@url \@href}%
\providecommand \@href[1]{\@@startlink{#1}\@@href}%
\providecommand \@@href[1]{\endgroup#1\@@endlink}%
\providecommand \@sanitize@url [0]{\catcode `\\12\catcode `\$12\catcode
  `\&12\catcode `\#12\catcode `\^12\catcode `\_12\catcode `\%12\relax}%
\providecommand \@@startlink[1]{}%
\providecommand \@@endlink[0]{}%
\providecommand \url  [0]{\begingroup\@sanitize@url \@url }%
\providecommand \@url [1]{\endgroup\@href {#1}{\urlprefix }}%
\providecommand \urlprefix  [0]{URL }%
\providecommand \Eprint [0]{\href }%
\providecommand \doibase [0]{http://dx.doi.org/}%
\providecommand \selectlanguage [0]{\@gobble}%
\providecommand \bibinfo  [0]{\@secondoftwo}%
\providecommand \bibfield  [0]{\@secondoftwo}%
\providecommand \translation [1]{[#1]}%
\providecommand \BibitemOpen [0]{}%
\providecommand \bibitemStop [0]{}%
\providecommand \bibitemNoStop [0]{.\EOS\space}%
\providecommand \EOS [0]{\spacefactor3000\relax}%
\providecommand \BibitemShut  [1]{\csname bibitem#1\endcsname}%
\let\auto@bib@innerbib\@empty
\bibitem [{\citenamefont {Nielsen}\ and\ \citenamefont
  {Chuang}(2000)}]{Nielsen00}%
  \BibitemOpen
  \bibfield  {author} {\bibinfo {author} {\bibfnamefont {M.~A.}\ \bibnamefont
  {Nielsen}}\ and\ \bibinfo {author} {\bibfnamefont {I.~L.}\ \bibnamefont
  {Chuang}},\ }\href@noop {} {\emph {\bibinfo {title} {Quantum {C}omputation
  and {Q}uantum {I}nformation}}}\ (\bibinfo  {publisher} {Cambridge University
  Press, Cambridge},\ \bibinfo {year} {2000})\BibitemShut {NoStop}%
\bibitem [{\citenamefont {Vitanov}\ \emph {et~al.}(2001)\citenamefont
  {Vitanov}, \citenamefont {Halfmann}, \citenamefont {Shore},\ and\
  \citenamefont {Bergmann}}]{physchem.52.1.763}%
  \BibitemOpen
  \bibfield  {author} {\bibinfo {author} {\bibfnamefont {N.~V.}\ \bibnamefont
  {Vitanov}}, \bibinfo {author} {\bibfnamefont {T.}~\bibnamefont {Halfmann}},
  \bibinfo {author} {\bibfnamefont {B.~W.}\ \bibnamefont {Shore}}, \ and\
  \bibinfo {author} {\bibfnamefont {K.}~\bibnamefont {Bergmann}},\ }\bibinfo
  {title} {Laser-induced population transfer by adiabatic passage techniques},\
  \href {\doibase 10.1146/annurev.physchem.52.1.763} {\bibfield  {journal}
  {\bibinfo  {journal} {Annu. Rev. Phys. Chem.}\ }\textbf {\bibinfo {volume}
  {52}},\ \bibinfo {pages} {763} (\bibinfo {year} {2001})}\BibitemShut
  {NoStop}%
\bibitem [{\citenamefont {Kr\'al}\ \emph {et~al.}(2007)\citenamefont {Kr\'al},
  \citenamefont {Thanopulos},\ and\ \citenamefont
  {Shapiro}}]{RevModPhys.79.53}%
  \BibitemOpen
  \bibfield  {author} {\bibinfo {author} {\bibfnamefont {P.}~\bibnamefont
  {Kr\'al}}, \bibinfo {author} {\bibfnamefont {I.}~\bibnamefont {Thanopulos}},
  \ and\ \bibinfo {author} {\bibfnamefont {M.}~\bibnamefont {Shapiro}},\
  }\bibinfo {title} {Colloquium: Coherently controlled adiabatic passage},\
  \href {\doibase 10.1103/RevModPhys.79.53} {\bibfield  {journal} {\bibinfo
  {journal} {Rev. Mod. Phys.}\ }\textbf {\bibinfo {volume} {79}},\ \bibinfo
  {pages} {53} (\bibinfo {year} {2007})}\BibitemShut {NoStop}%
\bibitem [{\citenamefont {Vitanov}\ \emph {et~al.}(2017)\citenamefont
  {Vitanov}, \citenamefont {Rangelov}, \citenamefont {Shore},\ and\
  \citenamefont {Bergmann}}]{RevModPhys.89.015006}%
  \BibitemOpen
  \bibfield  {author} {\bibinfo {author} {\bibfnamefont {N.~V.}\ \bibnamefont
  {Vitanov}}, \bibinfo {author} {\bibfnamefont {A.~A.}\ \bibnamefont
  {Rangelov}}, \bibinfo {author} {\bibfnamefont {B.~W.}\ \bibnamefont {Shore}},
  \ and\ \bibinfo {author} {\bibfnamefont {K.}~\bibnamefont {Bergmann}},\
  }\bibinfo {title} {Stimulated {R}aman adiabatic passage in physics,
  chemistry, and beyond},\ \href {\doibase 10.1103/RevModPhys.89.015006}
  {\bibfield  {journal} {\bibinfo  {journal} {Rev. Mod. Phys.}\ }\textbf
  {\bibinfo {volume} {89}},\ \bibinfo {pages} {015006} (\bibinfo {year}
  {2017})}\BibitemShut {NoStop}%
\bibitem [{\citenamefont {Daems}\ \emph {et~al.}(2013)\citenamefont {Daems},
  \citenamefont {Ruschhaupt}, \citenamefont {Sugny},\ and\ \citenamefont
  {Gu\'erin}}]{daems13}%
  \BibitemOpen
  \bibfield  {author} {\bibinfo {author} {\bibfnamefont {D.}~\bibnamefont
  {Daems}}, \bibinfo {author} {\bibfnamefont {A.}~\bibnamefont {Ruschhaupt}},
  \bibinfo {author} {\bibfnamefont {D.}~\bibnamefont {Sugny}}, \ and\ \bibinfo
  {author} {\bibfnamefont {S.}~\bibnamefont {Gu\'erin}},\ }\bibinfo {title}
  {Robust quantum control by a single-shot shaped pulse},\ \href {\doibase
  10.1103/PhysRevLett.111.050404} {\bibfield  {journal} {\bibinfo  {journal}
  {Phys. Rev. Lett.}\ }\textbf {\bibinfo {volume} {111}},\ \bibinfo {pages}
  {050404} (\bibinfo {year} {2013})}\BibitemShut {NoStop}%
\bibitem [{\citenamefont {Barnes}\ \emph {et~al.}(2015)\citenamefont {Barnes},
  \citenamefont {Wang},\ and\ \citenamefont {Sarma}}]{Barnes2015}%
  \BibitemOpen
  \bibfield  {author} {\bibinfo {author} {\bibfnamefont {E.}~\bibnamefont
  {Barnes}}, \bibinfo {author} {\bibfnamefont {X.}~\bibnamefont {Wang}}, \ and\
  \bibinfo {author} {\bibfnamefont {S.~D.}\ \bibnamefont {Sarma}},\ }\bibinfo
  {title} {Robust quantum control using smooth pulses and topological
  winding},\ \href {\doibase 10.1038/srep12685} {\bibfield  {journal} {\bibinfo
   {journal} {Scientific Reports}\ }\textbf {\bibinfo {volume} {5}},\ \bibinfo
  {pages} {12685} (\bibinfo {year} {2015})}\BibitemShut {NoStop}%
\bibitem [{\citenamefont {Van-Damme}\ \emph {et~al.}(2017)\citenamefont
  {Van-Damme}, \citenamefont {Schraft}, \citenamefont {Genov}, \citenamefont
  {Sugny}, \citenamefont {Halfmann},\ and\ \citenamefont
  {Gu\'erin}}]{PhysRevA.96.022309}%
  \BibitemOpen
  \bibfield  {author} {\bibinfo {author} {\bibfnamefont {L.}~\bibnamefont
  {Van-Damme}}, \bibinfo {author} {\bibfnamefont {D.}~\bibnamefont {Schraft}},
  \bibinfo {author} {\bibfnamefont {G.~T.}\ \bibnamefont {Genov}}, \bibinfo
  {author} {\bibfnamefont {D.}~\bibnamefont {Sugny}}, \bibinfo {author}
  {\bibfnamefont {T.}~\bibnamefont {Halfmann}}, \ and\ \bibinfo {author}
  {\bibfnamefont {S.}~\bibnamefont {Gu\'erin}},\ }\bibinfo {title} {Robust
  {NOT} gate by single-shot-shaped pulses: Demonstration of the efficiency of
  the pulses in rephasing atomic coherences},\ \href {\doibase
  10.1103/PhysRevA.96.022309} {\bibfield  {journal} {\bibinfo  {journal} {Phys.
  Rev. A}\ }\textbf {\bibinfo {volume} {96}},\ \bibinfo {pages} {022309}
  (\bibinfo {year} {2017})}\BibitemShut {NoStop}%
\bibitem [{\citenamefont {Zeng}\ \emph {et~al.}(2018)\citenamefont {Zeng},
  \citenamefont {Deng}, \citenamefont {Russo},\ and\ \citenamefont
  {Barnes}}]{Zeng2018}%
  \BibitemOpen
  \bibfield  {author} {\bibinfo {author} {\bibfnamefont {J.}~\bibnamefont
  {Zeng}}, \bibinfo {author} {\bibfnamefont {X.-H.}\ \bibnamefont {Deng}},
  \bibinfo {author} {\bibfnamefont {A.}~\bibnamefont {Russo}}, \ and\ \bibinfo
  {author} {\bibfnamefont {E.}~\bibnamefont {Barnes}},\ }\bibinfo {title}
  {General solution to inhomogeneous dephasing and smooth pulse dynamical
  decoupling},\ \href {\doibase 10.1088/1367-2630/aaafe9} {\bibfield  {journal}
  {\bibinfo  {journal} {New Journal of Physics}\ }\textbf {\bibinfo {volume}
  {20}},\ \bibinfo {pages} {033011} (\bibinfo {year} {2018})}\BibitemShut
  {NoStop}%
\bibitem [{\citenamefont {G\"ung\"ord\"u}\ and\ \citenamefont
  {Kestner}(2019)}]{Gngrd2019}%
  \BibitemOpen
  \bibfield  {author} {\bibinfo {author} {\bibfnamefont {U.}~\bibnamefont
  {G\"ung\"ord\"u}}\ and\ \bibinfo {author} {\bibfnamefont {J.~P.}\
  \bibnamefont {Kestner}},\ }\bibinfo {title} {Analytically parametrized
  solutions for robust quantum control using smooth pulses},\ \href {\doibase
  10.1103/PhysRevA.100.062310} {\bibfield  {journal} {\bibinfo  {journal}
  {Phys. Rev. A}\ }\textbf {\bibinfo {volume} {100}},\ \bibinfo {pages}
  {062310} (\bibinfo {year} {2019})}\BibitemShut {NoStop}%
\bibitem [{\citenamefont {Chen}\ \emph {et~al.}(2011)\citenamefont {Chen},
  \citenamefont {Torrontegui},\ and\ \citenamefont
  {Muga}}]{PhysRevA.83.062116}%
  \BibitemOpen
  \bibfield  {author} {\bibinfo {author} {\bibfnamefont {X.}~\bibnamefont
  {Chen}}, \bibinfo {author} {\bibfnamefont {E.}~\bibnamefont {Torrontegui}}, \
  and\ \bibinfo {author} {\bibfnamefont {J.~G.}\ \bibnamefont {Muga}},\
  }\bibinfo {title} {Lewis-{R}iesenfeld invariants and transitionless quantum
  driving},\ \href {\doibase 10.1103/PhysRevA.83.062116} {\bibfield  {journal}
  {\bibinfo  {journal} {Phys. Rev. A}\ }\textbf {\bibinfo {volume} {83}},\
  \bibinfo {pages} {062116} (\bibinfo {year} {2011})}\BibitemShut {NoStop}%
\bibitem [{\citenamefont {Ruschhaupt}\ \emph {et~al.}(2012)\citenamefont
  {Ruschhaupt}, \citenamefont {Chen}, \citenamefont {Alonso},\ and\
  \citenamefont {Muga}}]{Ruschhaupt2012}%
  \BibitemOpen
  \bibfield  {author} {\bibinfo {author} {\bibfnamefont {A.}~\bibnamefont
  {Ruschhaupt}}, \bibinfo {author} {\bibfnamefont {X.}~\bibnamefont {Chen}},
  \bibinfo {author} {\bibfnamefont {D.}~\bibnamefont {Alonso}}, \ and\ \bibinfo
  {author} {\bibfnamefont {J.~G.}\ \bibnamefont {Muga}},\ }\bibinfo {title}
  {Optimally robust shortcuts to population inversion in two-level quantum
  systems},\ \href {\doibase 10.1088/1367-2630/14/9/093040} {\bibfield
  {journal} {\bibinfo  {journal} {New Journal of Physics}\ }\textbf {\bibinfo
  {volume} {14}},\ \bibinfo {pages} {093040} (\bibinfo {year}
  {2012})}\BibitemShut {NoStop}%
\bibitem [{\citenamefont {Lu}\ \emph {et~al.}(2013)\citenamefont {Lu},
  \citenamefont {Chen}, \citenamefont {Ruschhaupt}, \citenamefont {Alonso},
  \citenamefont {Gu\'erin},\ and\ \citenamefont {Muga}}]{PhysRevA.88.033406}%
  \BibitemOpen
  \bibfield  {author} {\bibinfo {author} {\bibfnamefont {X.-J.}\ \bibnamefont
  {Lu}}, \bibinfo {author} {\bibfnamefont {X.}~\bibnamefont {Chen}}, \bibinfo
  {author} {\bibfnamefont {A.}~\bibnamefont {Ruschhaupt}}, \bibinfo {author}
  {\bibfnamefont {D.}~\bibnamefont {Alonso}}, \bibinfo {author} {\bibfnamefont
  {S.}~\bibnamefont {Gu\'erin}}, \ and\ \bibinfo {author} {\bibfnamefont
  {J.~G.}\ \bibnamefont {Muga}},\ }\bibinfo {title} {Fast and robust population
  transfer in two-level quantum systems with dephasing noise and/or systematic
  frequency errors},\ \href {\doibase 10.1103/PhysRevA.88.033406} {\bibfield
  {journal} {\bibinfo  {journal} {Phys. Rev. A}\ }\textbf {\bibinfo {volume}
  {88}},\ \bibinfo {pages} {033406} (\bibinfo {year} {2013})}\BibitemShut
  {NoStop}%
\bibitem [{\citenamefont {Laforgue}\ \emph {et~al.}(2019)\citenamefont
  {Laforgue}, \citenamefont {Chen},\ and\ \citenamefont
  {Gu\'erin}}]{PhysRevA.100.023415}%
  \BibitemOpen
  \bibfield  {author} {\bibinfo {author} {\bibfnamefont {X.}~\bibnamefont
  {Laforgue}}, \bibinfo {author} {\bibfnamefont {X.}~\bibnamefont {Chen}}, \
  and\ \bibinfo {author} {\bibfnamefont {S.}~\bibnamefont {Gu\'erin}},\
  }\bibinfo {title} {Robust stimulated {R}aman exact passage using shaped
  pulses},\ \href {\doibase 10.1103/PhysRevA.100.023415} {\bibfield  {journal}
  {\bibinfo  {journal} {Phys. Rev. A}\ }\textbf {\bibinfo {volume} {100}},\
  \bibinfo {pages} {023415} (\bibinfo {year} {2019})}\BibitemShut {NoStop}%
\bibitem [{\citenamefont {Song}\ \emph {et~al.}(2017)\citenamefont {Song},
  \citenamefont {Deng}, \citenamefont {Lamata},\ and\ \citenamefont
  {Muga}}]{PhysRevA.95.022332}%
  \BibitemOpen
  \bibfield  {author} {\bibinfo {author} {\bibfnamefont {X.-K.}\ \bibnamefont
  {Song}}, \bibinfo {author} {\bibfnamefont {F.-G.}\ \bibnamefont {Deng}},
  \bibinfo {author} {\bibfnamefont {L.}~\bibnamefont {Lamata}}, \ and\ \bibinfo
  {author} {\bibfnamefont {J.~G.}\ \bibnamefont {Muga}},\ }\bibinfo {title}
  {Robust state preparation in quantum simulations of {D}irac dynamics},\ \href
  {\doibase 10.1103/PhysRevA.95.022332} {\bibfield  {journal} {\bibinfo
  {journal} {Phys. Rev. A}\ }\textbf {\bibinfo {volume} {95}},\ \bibinfo
  {pages} {022332} (\bibinfo {year} {2017})}\BibitemShut {NoStop}%
\bibitem [{\citenamefont {Levy}\ \emph {et~al.}(2018)\citenamefont {Levy},
  \citenamefont {Kiely}, \citenamefont {Muga}, \citenamefont {Kosloff},\ and\
  \citenamefont {Torrontegui}}]{Levy2018}%
  \BibitemOpen
  \bibfield  {author} {\bibinfo {author} {\bibfnamefont {A.}~\bibnamefont
  {Levy}}, \bibinfo {author} {\bibfnamefont {A.}~\bibnamefont {Kiely}},
  \bibinfo {author} {\bibfnamefont {J.~G.}\ \bibnamefont {Muga}}, \bibinfo
  {author} {\bibfnamefont {R.}~\bibnamefont {Kosloff}}, \ and\ \bibinfo
  {author} {\bibfnamefont {E.}~\bibnamefont {Torrontegui}},\ }\bibinfo {title}
  {Noise resistant quantum control using dynamical invariants},\ \href
  {\doibase 10.1088/1367-2630/aaa9e5} {\bibfield  {journal} {\bibinfo
  {journal} {New Journal of Physics}\ }\textbf {\bibinfo {volume} {20}},\
  \bibinfo {pages} {025006} (\bibinfo {year} {2018})}\BibitemShut {NoStop}%
\bibitem [{\citenamefont {Yu}\ \emph {et~al.}(2018)\citenamefont {Yu},
  \citenamefont {Zhang}, \citenamefont {Ban},\ and\ \citenamefont
  {Chen}}]{PhysRevA.97.062317}%
  \BibitemOpen
  \bibfield  {author} {\bibinfo {author} {\bibfnamefont {X.-T.}\ \bibnamefont
  {Yu}}, \bibinfo {author} {\bibfnamefont {Q.}~\bibnamefont {Zhang}}, \bibinfo
  {author} {\bibfnamefont {Y.}~\bibnamefont {Ban}}, \ and\ \bibinfo {author}
  {\bibfnamefont {X.}~\bibnamefont {Chen}},\ }\bibinfo {title} {Fast and robust
  control of two interacting spins},\ \href {\doibase
  10.1103/PhysRevA.97.062317} {\bibfield  {journal} {\bibinfo  {journal} {Phys.
  Rev. A}\ }\textbf {\bibinfo {volume} {97}},\ \bibinfo {pages} {062317}
  (\bibinfo {year} {2018})}\BibitemShut {NoStop}%
\bibitem [{\citenamefont {Gu\'ery-Odelin}\ and\ \citenamefont
  {Muga}(2014)}]{PhysRevA.90.063425}%
  \BibitemOpen
  \bibfield  {author} {\bibinfo {author} {\bibfnamefont {D.}~\bibnamefont
  {Gu\'ery-Odelin}}\ and\ \bibinfo {author} {\bibfnamefont {J.~G.}\
  \bibnamefont {Muga}},\ }\bibinfo {title} {Transport in a harmonic trap:
  Shortcuts to adiabaticity and robust protocols},\ \href {\doibase
  10.1103/PhysRevA.90.063425} {\bibfield  {journal} {\bibinfo  {journal} {Phys.
  Rev. A}\ }\textbf {\bibinfo {volume} {90}},\ \bibinfo {pages} {063425}
  (\bibinfo {year} {2014})}\BibitemShut {NoStop}%
\bibitem [{\citenamefont {Kang}\ \emph {et~al.}(2020)\citenamefont {Kang},
  \citenamefont {Shi}, \citenamefont {Song},\ and\ \citenamefont
  {Xia}}]{kang20}%
  \BibitemOpen
  \bibfield  {author} {\bibinfo {author} {\bibfnamefont {Y.-H.}\ \bibnamefont
  {Kang}}, \bibinfo {author} {\bibfnamefont {Z.-C.}\ \bibnamefont {Shi}},
  \bibinfo {author} {\bibfnamefont {J.}~\bibnamefont {Song}}, \ and\ \bibinfo
  {author} {\bibfnamefont {Y.}~\bibnamefont {Xia}},\ }\bibinfo {title} {Robust
  generation of logical qubit singlet states with reverse engineering and
  optimal control with spin qubits},\ \href {https://arxiv.org/abs/2009.09411}
  {\bibfield  {journal} {\bibinfo  {journal} {arXiv: 2009.09411}\ } (\bibinfo
  {year} {2020})}\BibitemShut {NoStop}%
\bibitem [{\citenamefont {Zhang}\ and\ \citenamefont
  {Rabitz}(1994)}]{PhysRevA.49.2241}%
  \BibitemOpen
  \bibfield  {author} {\bibinfo {author} {\bibfnamefont {H.}~\bibnamefont
  {Zhang}}\ and\ \bibinfo {author} {\bibfnamefont {H.}~\bibnamefont {Rabitz}},\
  }\bibinfo {title} {Robust optimal control of quantum molecular systems in the
  presence of disturbances and uncertainties},\ \href {\doibase
  10.1103/PhysRevA.49.2241} {\bibfield  {journal} {\bibinfo  {journal} {Phys.
  Rev. A}\ }\textbf {\bibinfo {volume} {49}},\ \bibinfo {pages} {2241}
  (\bibinfo {year} {1994})}\BibitemShut {NoStop}%
\bibitem [{\citenamefont {Rabitz}(2002)}]{PhysRevA.66.063405}%
  \BibitemOpen
  \bibfield  {author} {\bibinfo {author} {\bibfnamefont {H.}~\bibnamefont
  {Rabitz}},\ }\bibinfo {title} {Optimal control of quantum systems: Origins of
  inherent robustness to control field fluctuations},\ \href {\doibase
  10.1103/PhysRevA.66.063405} {\bibfield  {journal} {\bibinfo  {journal} {Phys.
  Rev. A}\ }\textbf {\bibinfo {volume} {66}},\ \bibinfo {pages} {063405}
  (\bibinfo {year} {2002})}\BibitemShut {NoStop}%
\bibitem [{\citenamefont {Turinici}\ and\ \citenamefont
  {Rabitz}(2004)}]{PhysRevA.70.063412}%
  \BibitemOpen
  \bibfield  {author} {\bibinfo {author} {\bibfnamefont {G.}~\bibnamefont
  {Turinici}}\ and\ \bibinfo {author} {\bibfnamefont {H.}~\bibnamefont
  {Rabitz}},\ }\bibinfo {title} {Optimally controlling the internal dynamics of
  a randomly oriented ensemble of molecules},\ \href {\doibase
  10.1103/PhysRevA.70.063412} {\bibfield  {journal} {\bibinfo  {journal} {Phys.
  Rev. A}\ }\textbf {\bibinfo {volume} {70}},\ \bibinfo {pages} {063412}
  (\bibinfo {year} {2004})}\BibitemShut {NoStop}%
\bibitem [{\citenamefont {Wang}\ \emph {et~al.}(2010)\citenamefont {Wang},
  \citenamefont {Bayat}, \citenamefont {Schirmer},\ and\ \citenamefont
  {Bose}}]{PhysRevA.81.032312}%
  \BibitemOpen
  \bibfield  {author} {\bibinfo {author} {\bibfnamefont {X.}~\bibnamefont
  {Wang}}, \bibinfo {author} {\bibfnamefont {A.}~\bibnamefont {Bayat}},
  \bibinfo {author} {\bibfnamefont {S.~G.}\ \bibnamefont {Schirmer}}, \ and\
  \bibinfo {author} {\bibfnamefont {S.}~\bibnamefont {Bose}},\ }\bibinfo
  {title} {Robust entanglement in antiferromagnetic {H}eisenberg chains by
  single-spin optimal control},\ \href {\doibase 10.1103/PhysRevA.81.032312}
  {\bibfield  {journal} {\bibinfo  {journal} {Phys. Rev. A}\ }\textbf {\bibinfo
  {volume} {81}},\ \bibinfo {pages} {032312} (\bibinfo {year}
  {2010})}\BibitemShut {NoStop}%
\bibitem [{\citenamefont {Gorman}\ \emph {et~al.}(2012)\citenamefont {Gorman},
  \citenamefont {Young},\ and\ \citenamefont {Whaley}}]{PhysRevA.86.012317}%
  \BibitemOpen
  \bibfield  {author} {\bibinfo {author} {\bibfnamefont {D.~J.}\ \bibnamefont
  {Gorman}}, \bibinfo {author} {\bibfnamefont {K.~C.}\ \bibnamefont {Young}}, \
  and\ \bibinfo {author} {\bibfnamefont {K.~B.}\ \bibnamefont {Whaley}},\
  }\bibinfo {title} {Overcoming dephasing noise with robust optimal control},\
  \href {\doibase 10.1103/PhysRevA.86.012317} {\bibfield  {journal} {\bibinfo
  {journal} {Phys. Rev. A}\ }\textbf {\bibinfo {volume} {86}},\ \bibinfo
  {pages} {012317} (\bibinfo {year} {2012})}\BibitemShut {NoStop}%
\bibitem [{\citenamefont {Low}\ \emph {et~al.}(2014)\citenamefont {Low},
  \citenamefont {Yoder},\ and\ \citenamefont {Chuang}}]{PhysRevA.89.022341}%
  \BibitemOpen
  \bibfield  {author} {\bibinfo {author} {\bibfnamefont {G.~H.}\ \bibnamefont
  {Low}}, \bibinfo {author} {\bibfnamefont {T.~J.}\ \bibnamefont {Yoder}}, \
  and\ \bibinfo {author} {\bibfnamefont {I.~L.}\ \bibnamefont {Chuang}},\
  }\bibinfo {title} {Optimal arbitrarily accurate composite pulse sequences},\
  \href {\doibase 10.1103/PhysRevA.89.022341} {\bibfield  {journal} {\bibinfo
  {journal} {Phys. Rev. A}\ }\textbf {\bibinfo {volume} {89}},\ \bibinfo
  {pages} {022341} (\bibinfo {year} {2014})}\BibitemShut {NoStop}%
\bibitem [{\citenamefont {Hocker}\ \emph {et~al.}(2014)\citenamefont {Hocker},
  \citenamefont {Brif}, \citenamefont {Grace}, \citenamefont {Donovan},
  \citenamefont {Ho}, \citenamefont {Tibbetts}, \citenamefont {Wu},\ and\
  \citenamefont {Rabitz}}]{PhysRevA.90.062309}%
  \BibitemOpen
  \bibfield  {author} {\bibinfo {author} {\bibfnamefont {D.}~\bibnamefont
  {Hocker}}, \bibinfo {author} {\bibfnamefont {C.}~\bibnamefont {Brif}},
  \bibinfo {author} {\bibfnamefont {M.~D.}\ \bibnamefont {Grace}}, \bibinfo
  {author} {\bibfnamefont {A.}~\bibnamefont {Donovan}}, \bibinfo {author}
  {\bibfnamefont {T.-S.}\ \bibnamefont {Ho}}, \bibinfo {author} {\bibfnamefont
  {K.~M.}\ \bibnamefont {Tibbetts}}, \bibinfo {author} {\bibfnamefont
  {R.}~\bibnamefont {Wu}}, \ and\ \bibinfo {author} {\bibfnamefont
  {H.}~\bibnamefont {Rabitz}},\ }\bibinfo {title} {Characterization of control
  noise effects in optimal quantum unitary dynamics},\ \href {\doibase
  10.1103/PhysRevA.90.062309} {\bibfield  {journal} {\bibinfo  {journal} {Phys.
  Rev. A}\ }\textbf {\bibinfo {volume} {90}},\ \bibinfo {pages} {062309}
  (\bibinfo {year} {2014})}\BibitemShut {NoStop}%
\bibitem [{\citenamefont {N\"obauer}\ \emph {et~al.}(2015)\citenamefont
  {N\"obauer}, \citenamefont {Angerer}, \citenamefont {Bartels}, \citenamefont
  {Trupke}, \citenamefont {Rotter}, \citenamefont {Schmiedmayer}, \citenamefont
  {Mintert},\ and\ \citenamefont {Majer}}]{PhysRevLett.115.190801}%
  \BibitemOpen
  \bibfield  {author} {\bibinfo {author} {\bibfnamefont {T.}~\bibnamefont
  {N\"obauer}}, \bibinfo {author} {\bibfnamefont {A.}~\bibnamefont {Angerer}},
  \bibinfo {author} {\bibfnamefont {B.}~\bibnamefont {Bartels}}, \bibinfo
  {author} {\bibfnamefont {M.}~\bibnamefont {Trupke}}, \bibinfo {author}
  {\bibfnamefont {S.}~\bibnamefont {Rotter}}, \bibinfo {author} {\bibfnamefont
  {J.}~\bibnamefont {Schmiedmayer}}, \bibinfo {author} {\bibfnamefont
  {F.}~\bibnamefont {Mintert}}, \ and\ \bibinfo {author} {\bibfnamefont
  {J.}~\bibnamefont {Majer}},\ }\bibinfo {title} {Smooth optimal quantum
  control for robust solid-state spin magnetometry},\ \href {\doibase
  10.1103/PhysRevLett.115.190801} {\bibfield  {journal} {\bibinfo  {journal}
  {Phys. Rev. Lett.}\ }\textbf {\bibinfo {volume} {115}},\ \bibinfo {pages}
  {190801} (\bibinfo {year} {2015})}\BibitemShut {NoStop}%
\bibitem [{\citenamefont {Glaser}\ \emph {et~al.}(2015)\citenamefont {Glaser},
  \citenamefont {Boscain}, \citenamefont {Calarco}, \citenamefont {Koch},
  \citenamefont {K\"{o}ckenberger}, \citenamefont {Kosloff}, \citenamefont
  {Kuprov}, \citenamefont {Luy}, \citenamefont {Schirmer}, \citenamefont
  {Schulte-Herbr\"{u}ggen}, \citenamefont {Sugny},\ and\ \citenamefont
  {Wilhelm}}]{Glaser2015}%
  \BibitemOpen
  \bibfield  {author} {\bibinfo {author} {\bibfnamefont {S.~J.}\ \bibnamefont
  {Glaser}}, \bibinfo {author} {\bibfnamefont {U.}~\bibnamefont {Boscain}},
  \bibinfo {author} {\bibfnamefont {T.}~\bibnamefont {Calarco}}, \bibinfo
  {author} {\bibfnamefont {C.~P.}\ \bibnamefont {Koch}}, \bibinfo {author}
  {\bibfnamefont {W.}~\bibnamefont {K\"{o}ckenberger}}, \bibinfo {author}
  {\bibfnamefont {R.}~\bibnamefont {Kosloff}}, \bibinfo {author} {\bibfnamefont
  {I.}~\bibnamefont {Kuprov}}, \bibinfo {author} {\bibfnamefont
  {B.}~\bibnamefont {Luy}}, \bibinfo {author} {\bibfnamefont {S.}~\bibnamefont
  {Schirmer}}, \bibinfo {author} {\bibfnamefont {T.}~\bibnamefont
  {Schulte-Herbr\"{u}ggen}}, \bibinfo {author} {\bibfnamefont {D.}~\bibnamefont
  {Sugny}}, \ and\ \bibinfo {author} {\bibfnamefont {F.~K.}\ \bibnamefont
  {Wilhelm}},\ }\bibinfo {title} {Training {S}chr\"{o}dinger's cat: quantum
  optimal control},\ \href {\doibase 10.1140/epjd/e2015-60464-1} {\bibfield
  {journal} {\bibinfo  {journal} {The European Physical Journal D}\ }\textbf
  {\bibinfo {volume} {69}},\ \bibinfo {pages} {279} (\bibinfo {year}
  {2015})}\BibitemShut {NoStop}%
\bibitem [{\citenamefont {Ass\'emat}\ \emph {et~al.}(2010)\citenamefont
  {Ass\'emat}, \citenamefont {Lapert}, \citenamefont {Zhang}, \citenamefont
  {Braun}, \citenamefont {Glaser},\ and\ \citenamefont
  {Sugny}}]{PhysRevA.82.013415}%
  \BibitemOpen
  \bibfield  {author} {\bibinfo {author} {\bibfnamefont {E.}~\bibnamefont
  {Ass\'emat}}, \bibinfo {author} {\bibfnamefont {M.}~\bibnamefont {Lapert}},
  \bibinfo {author} {\bibfnamefont {Y.}~\bibnamefont {Zhang}}, \bibinfo
  {author} {\bibfnamefont {M.}~\bibnamefont {Braun}}, \bibinfo {author}
  {\bibfnamefont {S.~J.}\ \bibnamefont {Glaser}}, \ and\ \bibinfo {author}
  {\bibfnamefont {D.}~\bibnamefont {Sugny}},\ }\bibinfo {title} {Simultaneous
  time-optimal control of the inversion of two spin-$\frac{1}{2}$ particles},\
  \href {\doibase 10.1103/PhysRevA.82.013415} {\bibfield  {journal} {\bibinfo
  {journal} {Phys. Rev. A}\ }\textbf {\bibinfo {volume} {82}},\ \bibinfo
  {pages} {013415} (\bibinfo {year} {2010})}\BibitemShut {NoStop}%
\bibitem [{\citenamefont {Dong}\ \emph {et~al.}(2016)\citenamefont {Dong},
  \citenamefont {Wu}, \citenamefont {Chen}, \citenamefont {Qi}, \citenamefont
  {Petersen},\ and\ \citenamefont {Nori}}]{Dong2016}%
  \BibitemOpen
  \bibfield  {author} {\bibinfo {author} {\bibfnamefont {D.}~\bibnamefont
  {Dong}}, \bibinfo {author} {\bibfnamefont {C.}~\bibnamefont {Wu}}, \bibinfo
  {author} {\bibfnamefont {C.}~\bibnamefont {Chen}}, \bibinfo {author}
  {\bibfnamefont {B.}~\bibnamefont {Qi}}, \bibinfo {author} {\bibfnamefont
  {I.~R.}\ \bibnamefont {Petersen}}, \ and\ \bibinfo {author} {\bibfnamefont
  {F.}~\bibnamefont {Nori}},\ }\bibinfo {title} {Learning robust pulses for
  generating universal quantum gates},\ \href {\doibase 10.1038/srep36090}
  {\bibfield  {journal} {\bibinfo  {journal} {Scientific Reports}\ }\textbf
  {\bibinfo {volume} {6}},\ \bibinfo {pages} {36090} (\bibinfo {year}
  {2016})}\BibitemShut {NoStop}%
\bibitem [{\citenamefont {Van~Damme}\ \emph {et~al.}(2017)\citenamefont
  {Van~Damme}, \citenamefont {Ansel}, \citenamefont {Glaser},\ and\
  \citenamefont {Sugny}}]{PhysRevA.95.063403}%
  \BibitemOpen
  \bibfield  {author} {\bibinfo {author} {\bibfnamefont {L.}~\bibnamefont
  {Van~Damme}}, \bibinfo {author} {\bibfnamefont {Q.}~\bibnamefont {Ansel}},
  \bibinfo {author} {\bibfnamefont {S.~J.}\ \bibnamefont {Glaser}}, \ and\
  \bibinfo {author} {\bibfnamefont {D.}~\bibnamefont {Sugny}},\ }\bibinfo
  {title} {Robust optimal control of two-level quantum systems},\ \href
  {\doibase 10.1103/PhysRevA.95.063403} {\bibfield  {journal} {\bibinfo
  {journal} {Phys. Rev. A}\ }\textbf {\bibinfo {volume} {95}},\ \bibinfo
  {pages} {063403} (\bibinfo {year} {2017})}\BibitemShut {NoStop}%
\bibitem [{\citenamefont {Wu}\ \emph {et~al.}(2019)\citenamefont {Wu},
  \citenamefont {Ding}, \citenamefont {Dong},\ and\ \citenamefont
  {Wang}}]{PhysRevA.99.042327}%
  \BibitemOpen
  \bibfield  {author} {\bibinfo {author} {\bibfnamefont {R.-B.}\ \bibnamefont
  {Wu}}, \bibinfo {author} {\bibfnamefont {H.}~\bibnamefont {Ding}}, \bibinfo
  {author} {\bibfnamefont {D.}~\bibnamefont {Dong}}, \ and\ \bibinfo {author}
  {\bibfnamefont {X.}~\bibnamefont {Wang}},\ }\bibinfo {title} {Learning robust
  and high-precision quantum controls},\ \href {\doibase
  10.1103/PhysRevA.99.042327} {\bibfield  {journal} {\bibinfo  {journal} {Phys.
  Rev. A}\ }\textbf {\bibinfo {volume} {99}},\ \bibinfo {pages} {042327}
  (\bibinfo {year} {2019})}\BibitemShut {NoStop}%
\bibitem [{\citenamefont {Tian}\ \emph {et~al.}(2020)\citenamefont {Tian},
  \citenamefont {Liu}, \citenamefont {Liu}, \citenamefont {Yang}, \citenamefont
  {Betzholz}, \citenamefont {Said}, \citenamefont {Jelezko},\ and\
  \citenamefont {Cai}}]{PhysRevA.102.043707}%
  \BibitemOpen
  \bibfield  {author} {\bibinfo {author} {\bibfnamefont {J.}~\bibnamefont
  {Tian}}, \bibinfo {author} {\bibfnamefont {H.}~\bibnamefont {Liu}}, \bibinfo
  {author} {\bibfnamefont {Y.}~\bibnamefont {Liu}}, \bibinfo {author}
  {\bibfnamefont {P.}~\bibnamefont {Yang}}, \bibinfo {author} {\bibfnamefont
  {R.}~\bibnamefont {Betzholz}}, \bibinfo {author} {\bibfnamefont {R.~S.}\
  \bibnamefont {Said}}, \bibinfo {author} {\bibfnamefont {F.}~\bibnamefont
  {Jelezko}}, \ and\ \bibinfo {author} {\bibfnamefont {J.}~\bibnamefont
  {Cai}},\ }\bibinfo {title} {Quantum optimal control using phase-modulated
  driving fields},\ \href {\doibase 10.1103/PhysRevA.102.043707} {\bibfield
  {journal} {\bibinfo  {journal} {Phys. Rev. A}\ }\textbf {\bibinfo {volume}
  {102}},\ \bibinfo {pages} {043707} (\bibinfo {year} {2020})}\BibitemShut
  {NoStop}%
\bibitem [{\citenamefont {Abraham}(1961)}]{Abraham61}%
  \BibitemOpen
  \bibfield  {author} {\bibinfo {author} {\bibfnamefont {A.}~\bibnamefont
  {Abraham}},\ }\href@noop {} {\emph {\bibinfo {title} {The {P}rinciples of
  {N}uclear {M}agnetism}}}\ (\bibinfo  {publisher} {Clarendon, Oxford},\
  \bibinfo {year} {1961})\BibitemShut {NoStop}%
\bibitem [{\citenamefont {Slichter}(1990)}]{Slichter90}%
  \BibitemOpen
  \bibfield  {author} {\bibinfo {author} {\bibfnamefont {C.~P.}\ \bibnamefont
  {Slichter}},\ }\href@noop {} {\emph {\bibinfo {title} {Principles of
  {M}agnetic {R}esonance}}}\ (\bibinfo  {publisher} {Springer, Berlin},\
  \bibinfo {year} {1990})\BibitemShut {NoStop}%
\bibitem [{\citenamefont {Freeman}(1997)}]{Freeman97}%
  \BibitemOpen
  \bibfield  {author} {\bibinfo {author} {\bibfnamefont {R.}~\bibnamefont
  {Freeman}},\ }\href@noop {} {\emph {\bibinfo {title} {Spin {C}horeography}}}\
  (\bibinfo  {publisher} {Spektrum, Oxford},\ \bibinfo {year}
  {1997})\BibitemShut {NoStop}%
\bibitem [{\citenamefont {Wang}\ \emph {et~al.}(2012)\citenamefont {Wang},
  \citenamefont {Bishop}, \citenamefont {Kestner}, \citenamefont {Barnes},
  \citenamefont {Sun},\ and\ \citenamefont {Sarma}}]{Wang2012}%
  \BibitemOpen
  \bibfield  {author} {\bibinfo {author} {\bibfnamefont {X.}~\bibnamefont
  {Wang}}, \bibinfo {author} {\bibfnamefont {L.~S.}\ \bibnamefont {Bishop}},
  \bibinfo {author} {\bibfnamefont {J.}~\bibnamefont {Kestner}}, \bibinfo
  {author} {\bibfnamefont {E.}~\bibnamefont {Barnes}}, \bibinfo {author}
  {\bibfnamefont {K.}~\bibnamefont {Sun}}, \ and\ \bibinfo {author}
  {\bibfnamefont {S.~D.}\ \bibnamefont {Sarma}},\ }\bibinfo {title} {Composite
  pulses for robust universal control of singlet{\textendash}triplet qubits},\
  \href {\doibase 10.1038/ncomms2003} {\bibfield  {journal} {\bibinfo
  {journal} {Nature Communications}\ }\textbf {\bibinfo {volume} {3}},\
  \bibinfo {pages} {997} (\bibinfo {year} {2012})}\BibitemShut {NoStop}%
\bibitem [{\citenamefont {Wang}\ \emph {et~al.}(2014)\citenamefont {Wang},
  \citenamefont {Bishop}, \citenamefont {Barnes}, \citenamefont {Kestner},\
  and\ \citenamefont {Sarma}}]{PhysRevA.89.022310}%
  \BibitemOpen
  \bibfield  {author} {\bibinfo {author} {\bibfnamefont {X.}~\bibnamefont
  {Wang}}, \bibinfo {author} {\bibfnamefont {L.~S.}\ \bibnamefont {Bishop}},
  \bibinfo {author} {\bibfnamefont {E.}~\bibnamefont {Barnes}}, \bibinfo
  {author} {\bibfnamefont {J.~P.}\ \bibnamefont {Kestner}}, \ and\ \bibinfo
  {author} {\bibfnamefont {S.~D.}\ \bibnamefont {Sarma}},\ }\bibinfo {title}
  {Robust quantum gates for singlet-triplet spin qubits using composite
  pulses},\ \href {\doibase 10.1103/PhysRevA.89.022310} {\bibfield  {journal}
  {\bibinfo  {journal} {Phys. Rev. A}\ }\textbf {\bibinfo {volume} {89}},\
  \bibinfo {pages} {022310} (\bibinfo {year} {2014})}\BibitemShut {NoStop}%
\bibitem [{\citenamefont {Kestner}\ \emph {et~al.}(2013)\citenamefont
  {Kestner}, \citenamefont {Wang}, \citenamefont {Bishop}, \citenamefont
  {Barnes},\ and\ \citenamefont {Das~Sarma}}]{PhysRevLett.110.140502}%
  \BibitemOpen
  \bibfield  {author} {\bibinfo {author} {\bibfnamefont {J.~P.}\ \bibnamefont
  {Kestner}}, \bibinfo {author} {\bibfnamefont {X.}~\bibnamefont {Wang}},
  \bibinfo {author} {\bibfnamefont {L.~S.}\ \bibnamefont {Bishop}}, \bibinfo
  {author} {\bibfnamefont {E.}~\bibnamefont {Barnes}}, \ and\ \bibinfo {author}
  {\bibfnamefont {S.}~\bibnamefont {Das~Sarma}},\ }\bibinfo {title}
  {Noise-resistant control for a spin qubit array},\ \href {\doibase
  10.1103/PhysRevLett.110.140502} {\bibfield  {journal} {\bibinfo  {journal}
  {Phys. Rev. Lett.}\ }\textbf {\bibinfo {volume} {110}},\ \bibinfo {pages}
  {140502} (\bibinfo {year} {2013})}\BibitemShut {NoStop}%
\bibitem [{\citenamefont {Yang}\ \emph {et~al.}(2018)\citenamefont {Yang},
  \citenamefont {Yung},\ and\ \citenamefont {Wang}}]{PhysRevA.97.042324}%
  \BibitemOpen
  \bibfield  {author} {\bibinfo {author} {\bibfnamefont {X.-C.}\ \bibnamefont
  {Yang}}, \bibinfo {author} {\bibfnamefont {M.-H.}\ \bibnamefont {Yung}}, \
  and\ \bibinfo {author} {\bibfnamefont {X.}~\bibnamefont {Wang}},\ }\bibinfo
  {title} {Neural-network-designed pulse sequences for robust control of
  singlet-triplet qubits},\ \href {\doibase 10.1103/PhysRevA.97.042324}
  {\bibfield  {journal} {\bibinfo  {journal} {Phys. Rev. A}\ }\textbf {\bibinfo
  {volume} {97}},\ \bibinfo {pages} {042324} (\bibinfo {year}
  {2018})}\BibitemShut {NoStop}%
\bibitem [{\citenamefont {Torosov}\ and\ \citenamefont
  {Vitanov}(2019{\natexlab{a}})}]{PhysRevA.99.013424}%
  \BibitemOpen
  \bibfield  {author} {\bibinfo {author} {\bibfnamefont {B.~T.}\ \bibnamefont
  {Torosov}}\ and\ \bibinfo {author} {\bibfnamefont {N.~V.}\ \bibnamefont
  {Vitanov}},\ }\bibinfo {title} {Robust high-fidelity coherent control of
  two-state systems by detuning pulses},\ \href {\doibase
  10.1103/PhysRevA.99.013424} {\bibfield  {journal} {\bibinfo  {journal} {Phys.
  Rev. A}\ }\textbf {\bibinfo {volume} {99}},\ \bibinfo {pages} {013424}
  (\bibinfo {year} {2019}{\natexlab{a}})}\BibitemShut {NoStop}%
\bibitem [{\citenamefont {Kyoseva}\ \emph {et~al.}(2019)\citenamefont
  {Kyoseva}, \citenamefont {Greener},\ and\ \citenamefont
  {Suchowski}}]{PhysRevA.100.032333}%
  \BibitemOpen
  \bibfield  {author} {\bibinfo {author} {\bibfnamefont {E.}~\bibnamefont
  {Kyoseva}}, \bibinfo {author} {\bibfnamefont {H.}~\bibnamefont {Greener}}, \
  and\ \bibinfo {author} {\bibfnamefont {H.}~\bibnamefont {Suchowski}},\
  }\bibinfo {title} {Detuning-modulated composite pulses for high-fidelity
  robust quantum control},\ \href {\doibase 10.1103/PhysRevA.100.032333}
  {\bibfield  {journal} {\bibinfo  {journal} {Phys. Rev. A}\ }\textbf {\bibinfo
  {volume} {100}},\ \bibinfo {pages} {032333} (\bibinfo {year}
  {2019})}\BibitemShut {NoStop}%
\bibitem [{\citenamefont {Genov}\ \emph {et~al.}(2014)\citenamefont {Genov},
  \citenamefont {Schraft}, \citenamefont {Halfmann},\ and\ \citenamefont
  {Vitanov}}]{PhysRevLett.113.043001}%
  \BibitemOpen
  \bibfield  {author} {\bibinfo {author} {\bibfnamefont {G.~T.}\ \bibnamefont
  {Genov}}, \bibinfo {author} {\bibfnamefont {D.}~\bibnamefont {Schraft}},
  \bibinfo {author} {\bibfnamefont {T.}~\bibnamefont {Halfmann}}, \ and\
  \bibinfo {author} {\bibfnamefont {N.~V.}\ \bibnamefont {Vitanov}},\ }\bibinfo
  {title} {Correction of arbitrary field errors in population inversion of
  quantum systems by universal composite pulses},\ \href {\doibase
  10.1103/PhysRevLett.113.043001} {\bibfield  {journal} {\bibinfo  {journal}
  {Phys. Rev. Lett.}\ }\textbf {\bibinfo {volume} {113}},\ \bibinfo {pages}
  {043001} (\bibinfo {year} {2014})}\BibitemShut {NoStop}%
\bibitem [{\citenamefont {Vitanov}(2011)}]{PhysRevA.84.065404}%
  \BibitemOpen
  \bibfield  {author} {\bibinfo {author} {\bibfnamefont {N.~V.}\ \bibnamefont
  {Vitanov}},\ }\bibinfo {title} {Arbitrarily accurate narrowband composite
  pulse sequences},\ \href {\doibase 10.1103/PhysRevA.84.065404} {\bibfield
  {journal} {\bibinfo  {journal} {Phys. Rev. A}\ }\textbf {\bibinfo {volume}
  {84}},\ \bibinfo {pages} {065404} (\bibinfo {year} {2011})}\BibitemShut
  {NoStop}%
\bibitem [{\citenamefont {Kyoseva}\ and\ \citenamefont
  {Vitanov}(2013)}]{PhysRevA.88.063410}%
  \BibitemOpen
  \bibfield  {author} {\bibinfo {author} {\bibfnamefont {E.}~\bibnamefont
  {Kyoseva}}\ and\ \bibinfo {author} {\bibfnamefont {N.~V.}\ \bibnamefont
  {Vitanov}},\ }\bibinfo {title} {Arbitrarily accurate passband composite
  pulses for dynamical suppression of amplitude noise},\ \href {\doibase
  10.1103/PhysRevA.88.063410} {\bibfield  {journal} {\bibinfo  {journal} {Phys.
  Rev. A}\ }\textbf {\bibinfo {volume} {88}},\ \bibinfo {pages} {063410}
  (\bibinfo {year} {2013})}\BibitemShut {NoStop}%
\bibitem [{\citenamefont {Torosov}\ and\ \citenamefont
  {Vitanov}(2019{\natexlab{b}})}]{PhysRevA.99.013402}%
  \BibitemOpen
  \bibfield  {author} {\bibinfo {author} {\bibfnamefont {B.~T.}\ \bibnamefont
  {Torosov}}\ and\ \bibinfo {author} {\bibfnamefont {N.~V.}\ \bibnamefont
  {Vitanov}},\ }\bibinfo {title} {Arbitrarily accurate variable rotations on
  the {B}loch sphere by composite pulse sequences},\ \href {\doibase
  10.1103/PhysRevA.99.013402} {\bibfield  {journal} {\bibinfo  {journal} {Phys.
  Rev. A}\ }\textbf {\bibinfo {volume} {99}},\ \bibinfo {pages} {013402}
  (\bibinfo {year} {2019}{\natexlab{b}})}\BibitemShut {NoStop}%
\bibitem [{\citenamefont {Dridi}\ \emph {et~al.}(2020)\citenamefont {Dridi},
  \citenamefont {Mejatty}, \citenamefont {Glaser},\ and\ \citenamefont
  {Sugny}}]{PhysRevA.101.012321}%
  \BibitemOpen
  \bibfield  {author} {\bibinfo {author} {\bibfnamefont {G.}~\bibnamefont
  {Dridi}}, \bibinfo {author} {\bibfnamefont {M.}~\bibnamefont {Mejatty}},
  \bibinfo {author} {\bibfnamefont {S.~J.}\ \bibnamefont {Glaser}}, \ and\
  \bibinfo {author} {\bibfnamefont {D.}~\bibnamefont {Sugny}},\ }\bibinfo
  {title} {Robust control of a {NOT} gate by composite pulses},\ \href
  {\doibase 10.1103/PhysRevA.101.012321} {\bibfield  {journal} {\bibinfo
  {journal} {Phys. Rev. A}\ }\textbf {\bibinfo {volume} {101}},\ \bibinfo
  {pages} {012321} (\bibinfo {year} {2020})}\BibitemShut {NoStop}%
\bibitem [{\citenamefont {Genov}\ \emph {et~al.}(2020)\citenamefont {Genov},
  \citenamefont {Hain}, \citenamefont {Vitanov},\ and\ \citenamefont
  {Halfmann}}]{PhysRevA.101.013827}%
  \BibitemOpen
  \bibfield  {author} {\bibinfo {author} {\bibfnamefont {G.~T.}\ \bibnamefont
  {Genov}}, \bibinfo {author} {\bibfnamefont {M.}~\bibnamefont {Hain}},
  \bibinfo {author} {\bibfnamefont {N.~V.}\ \bibnamefont {Vitanov}}, \ and\
  \bibinfo {author} {\bibfnamefont {T.}~\bibnamefont {Halfmann}},\ }\bibinfo
  {title} {Universal composite pulses for efficient population inversion with
  an arbitrary excitation profile},\ \href {\doibase
  10.1103/PhysRevA.101.013827} {\bibfield  {journal} {\bibinfo  {journal}
  {Phys. Rev. A}\ }\textbf {\bibinfo {volume} {101}},\ \bibinfo {pages}
  {013827} (\bibinfo {year} {2020})}\BibitemShut {NoStop}%
\bibitem [{\citenamefont {Tomita}\ \emph {et~al.}(2010)\citenamefont {Tomita},
  \citenamefont {Merrill},\ and\ \citenamefont {Brown}}]{Tomita2010}%
  \BibitemOpen
  \bibfield  {author} {\bibinfo {author} {\bibfnamefont {Y.}~\bibnamefont
  {Tomita}}, \bibinfo {author} {\bibfnamefont {J.~T.}\ \bibnamefont {Merrill}},
  \ and\ \bibinfo {author} {\bibfnamefont {K.~R.}\ \bibnamefont {Brown}},\
  }\bibinfo {title} {Multi-qubit compensation sequences},\ \href {\doibase
  10.1088/1367-2630/12/1/015002} {\bibfield  {journal} {\bibinfo  {journal}
  {New Journal of Physics}\ }\textbf {\bibinfo {volume} {12}},\ \bibinfo
  {pages} {015002} (\bibinfo {year} {2010})}\BibitemShut {NoStop}%
\bibitem [{\citenamefont {Dunning}\ \emph {et~al.}(2014)\citenamefont
  {Dunning}, \citenamefont {Gregory}, \citenamefont {Bateman}, \citenamefont
  {Cooper}, \citenamefont {Himsworth}, \citenamefont {Jones},\ and\
  \citenamefont {Freegarde}}]{PhysRevA.90.033608}%
  \BibitemOpen
  \bibfield  {author} {\bibinfo {author} {\bibfnamefont {A.}~\bibnamefont
  {Dunning}}, \bibinfo {author} {\bibfnamefont {R.}~\bibnamefont {Gregory}},
  \bibinfo {author} {\bibfnamefont {J.}~\bibnamefont {Bateman}}, \bibinfo
  {author} {\bibfnamefont {N.}~\bibnamefont {Cooper}}, \bibinfo {author}
  {\bibfnamefont {M.}~\bibnamefont {Himsworth}}, \bibinfo {author}
  {\bibfnamefont {J.~A.}\ \bibnamefont {Jones}}, \ and\ \bibinfo {author}
  {\bibfnamefont {T.}~\bibnamefont {Freegarde}},\ }\bibinfo {title} {Composite
  pulses for interferometry in a thermal cold atom cloud},\ \href {\doibase
  10.1103/PhysRevA.90.033608} {\bibfield  {journal} {\bibinfo  {journal} {Phys.
  Rev. A}\ }\textbf {\bibinfo {volume} {90}},\ \bibinfo {pages} {033608}
  (\bibinfo {year} {2014})}\BibitemShut {NoStop}%
\bibitem [{\citenamefont {Merrill}\ \emph {et~al.}(2014)\citenamefont
  {Merrill}, \citenamefont {Doret}, \citenamefont {Vittorini}, \citenamefont
  {Addison},\ and\ \citenamefont {Brown}}]{PhysRevA.90.040301}%
  \BibitemOpen
  \bibfield  {author} {\bibinfo {author} {\bibfnamefont {J.~T.}\ \bibnamefont
  {Merrill}}, \bibinfo {author} {\bibfnamefont {S.~C.}\ \bibnamefont {Doret}},
  \bibinfo {author} {\bibfnamefont {G.}~\bibnamefont {Vittorini}}, \bibinfo
  {author} {\bibfnamefont {J.~P.}\ \bibnamefont {Addison}}, \ and\ \bibinfo
  {author} {\bibfnamefont {K.~R.}\ \bibnamefont {Brown}},\ }\bibinfo {title}
  {Transformed composite sequences for improved qubit addressing},\ \href
  {\doibase 10.1103/PhysRevA.90.040301} {\bibfield  {journal} {\bibinfo
  {journal} {Phys. Rev. A}\ }\textbf {\bibinfo {volume} {90}},\ \bibinfo
  {pages} {040301} (\bibinfo {year} {2014})}\BibitemShut {NoStop}%
\bibitem [{\citenamefont {Cohen}\ \emph {et~al.}(2016)\citenamefont {Cohen},
  \citenamefont {Rotem},\ and\ \citenamefont {Retzker}}]{PhysRevA.93.032340}%
  \BibitemOpen
  \bibfield  {author} {\bibinfo {author} {\bibfnamefont {I.}~\bibnamefont
  {Cohen}}, \bibinfo {author} {\bibfnamefont {A.}~\bibnamefont {Rotem}}, \ and\
  \bibinfo {author} {\bibfnamefont {A.}~\bibnamefont {Retzker}},\ }\bibinfo
  {title} {Refocusing two-qubit-gate noise for trapped ions by composite
  pulses},\ \href {\doibase 10.1103/PhysRevA.93.032340} {\bibfield  {journal}
  {\bibinfo  {journal} {Phys. Rev. A}\ }\textbf {\bibinfo {volume} {93}},\
  \bibinfo {pages} {032340} (\bibinfo {year} {2016})}\BibitemShut {NoStop}%
\bibitem [{\citenamefont {Ivanov}\ \emph {et~al.}(2013)\citenamefont {Ivanov},
  \citenamefont {Vitanov},\ and\ \citenamefont {Korolkova}}]{Ivanov2013}%
  \BibitemOpen
  \bibfield  {author} {\bibinfo {author} {\bibfnamefont {S.~S.}\ \bibnamefont
  {Ivanov}}, \bibinfo {author} {\bibfnamefont {N.~V.}\ \bibnamefont {Vitanov}},
  \ and\ \bibinfo {author} {\bibfnamefont {N.~V.}\ \bibnamefont {Korolkova}},\
  }\bibinfo {title} {Creation of arbitrary {D}icke and {NOON} states of
  trapped-ion qubits by global addressing with composite pulses},\ \href
  {\doibase 10.1088/1367-2630/15/2/023039} {\bibfield  {journal} {\bibinfo
  {journal} {New Journal of Physics}\ }\textbf {\bibinfo {volume} {15}},\
  \bibinfo {pages} {023039} (\bibinfo {year} {2013})}\BibitemShut {NoStop}%
\bibitem [{\citenamefont {Calderon-Vargas}\ and\ \citenamefont
  {Kestner}(2017)}]{PhysRevLett.118.150502}%
  \BibitemOpen
  \bibfield  {author} {\bibinfo {author} {\bibfnamefont {F.~A.}\ \bibnamefont
  {Calderon-Vargas}}\ and\ \bibinfo {author} {\bibfnamefont {J.~P.}\
  \bibnamefont {Kestner}},\ }\bibinfo {title} {Dynamically correcting a
  $\mathrm{CNOT}$ gate for any systematic logical error},\ \href {\doibase
  10.1103/PhysRevLett.118.150502} {\bibfield  {journal} {\bibinfo  {journal}
  {Phys. Rev. Lett.}\ }\textbf {\bibinfo {volume} {118}},\ \bibinfo {pages}
  {150502} (\bibinfo {year} {2017})}\BibitemShut {NoStop}%
\bibitem [{\citenamefont {Brown}\ \emph {et~al.}(2004)\citenamefont {Brown},
  \citenamefont {Harrow},\ and\ \citenamefont {Chuang}}]{PhysRevA.70.052318}%
  \BibitemOpen
  \bibfield  {author} {\bibinfo {author} {\bibfnamefont {K.~R.}\ \bibnamefont
  {Brown}}, \bibinfo {author} {\bibfnamefont {A.~W.}\ \bibnamefont {Harrow}}, \
  and\ \bibinfo {author} {\bibfnamefont {I.~L.}\ \bibnamefont {Chuang}},\
  }\bibinfo {title} {Arbitrarily accurate composite pulse sequences},\ \href
  {\doibase 10.1103/PhysRevA.70.052318} {\bibfield  {journal} {\bibinfo
  {journal} {Phys. Rev. A}\ }\textbf {\bibinfo {volume} {70}},\ \bibinfo
  {pages} {052318} (\bibinfo {year} {2004})}\BibitemShut {NoStop}%
\bibitem [{\citenamefont {Torosov}\ \emph {et~al.}(2011)\citenamefont
  {Torosov}, \citenamefont {Gu\'erin},\ and\ \citenamefont
  {Vitanov}}]{PhysRevLett.106.233001}%
  \BibitemOpen
  \bibfield  {author} {\bibinfo {author} {\bibfnamefont {B.~T.}\ \bibnamefont
  {Torosov}}, \bibinfo {author} {\bibfnamefont {S.}~\bibnamefont {Gu\'erin}}, \
  and\ \bibinfo {author} {\bibfnamefont {N.~V.}\ \bibnamefont {Vitanov}},\
  }\bibinfo {title} {High-fidelity adiabatic passage by composite sequences of
  chirped pulses},\ \href {\doibase 10.1103/PhysRevLett.106.233001} {\bibfield
  {journal} {\bibinfo  {journal} {Phys. Rev. Lett.}\ }\textbf {\bibinfo
  {volume} {106}},\ \bibinfo {pages} {233001} (\bibinfo {year}
  {2011})}\BibitemShut {NoStop}%
\bibitem [{\citenamefont {Kabytayev}\ \emph {et~al.}(2014)\citenamefont
  {Kabytayev}, \citenamefont {Green}, \citenamefont {Khodjasteh}, \citenamefont
  {Biercuk}, \citenamefont {Viola},\ and\ \citenamefont
  {Brown}}]{PhysRevA.90.012316}%
  \BibitemOpen
  \bibfield  {author} {\bibinfo {author} {\bibfnamefont {C.}~\bibnamefont
  {Kabytayev}}, \bibinfo {author} {\bibfnamefont {T.~J.}\ \bibnamefont
  {Green}}, \bibinfo {author} {\bibfnamefont {K.}~\bibnamefont {Khodjasteh}},
  \bibinfo {author} {\bibfnamefont {M.~J.}\ \bibnamefont {Biercuk}}, \bibinfo
  {author} {\bibfnamefont {L.}~\bibnamefont {Viola}}, \ and\ \bibinfo {author}
  {\bibfnamefont {K.~R.}\ \bibnamefont {Brown}},\ }\bibinfo {title} {Robustness
  of composite pulses to time-dependent control noise},\ \href {\doibase
  10.1103/PhysRevA.90.012316} {\bibfield  {journal} {\bibinfo  {journal} {Phys.
  Rev. A}\ }\textbf {\bibinfo {volume} {90}},\ \bibinfo {pages} {012316}
  (\bibinfo {year} {2014})}\BibitemShut {NoStop}%
\bibitem [{\citenamefont {Jones}(2013)}]{PhysRevA.87.052317}%
  \BibitemOpen
  \bibfield  {author} {\bibinfo {author} {\bibfnamefont {J.~A.}\ \bibnamefont
  {Jones}},\ }\bibinfo {title} {Designing short robust {NOT} gates for quantum
  computation},\ \href {\doibase 10.1103/PhysRevA.87.052317} {\bibfield
  {journal} {\bibinfo  {journal} {Phys. Rev. A}\ }\textbf {\bibinfo {volume}
  {87}},\ \bibinfo {pages} {052317} (\bibinfo {year} {2013})}\BibitemShut
  {NoStop}%
\bibitem [{\citenamefont {Casanova}\ \emph {et~al.}(2015)\citenamefont
  {Casanova}, \citenamefont {Wang}, \citenamefont {Haase},\ and\ \citenamefont
  {Plenio}}]{PhysRevA.92.042304}%
  \BibitemOpen
  \bibfield  {author} {\bibinfo {author} {\bibfnamefont {J.}~\bibnamefont
  {Casanova}}, \bibinfo {author} {\bibfnamefont {Z.-Y.}\ \bibnamefont {Wang}},
  \bibinfo {author} {\bibfnamefont {J.~F.}\ \bibnamefont {Haase}}, \ and\
  \bibinfo {author} {\bibfnamefont {M.~B.}\ \bibnamefont {Plenio}},\ }\bibinfo
  {title} {Robust dynamical decoupling sequences for individual-nuclear-spin
  addressing},\ \href {\doibase 10.1103/PhysRevA.92.042304} {\bibfield
  {journal} {\bibinfo  {journal} {Phys. Rev. A}\ }\textbf {\bibinfo {volume}
  {92}},\ \bibinfo {pages} {042304} (\bibinfo {year} {2015})}\BibitemShut
  {NoStop}%
\bibitem [{\citenamefont {Demeter}(2016)}]{PhysRevA.93.023830}%
  \BibitemOpen
  \bibfield  {author} {\bibinfo {author} {\bibfnamefont {G.}~\bibnamefont
  {Demeter}},\ }\bibinfo {title} {Composite pulses for high-fidelity population
  inversion in optically dense, inhomogeneously broadened atomic ensembles},\
  \href {\doibase 10.1103/PhysRevA.93.023830} {\bibfield  {journal} {\bibinfo
  {journal} {Phys. Rev. A}\ }\textbf {\bibinfo {volume} {93}},\ \bibinfo
  {pages} {023830} (\bibinfo {year} {2016})}\BibitemShut {NoStop}%
\bibitem [{\citenamefont {Genov}\ \emph {et~al.}(2017)\citenamefont {Genov},
  \citenamefont {Schraft}, \citenamefont {Vitanov},\ and\ \citenamefont
  {Halfmann}}]{PhysRevLett.118.133202}%
  \BibitemOpen
  \bibfield  {author} {\bibinfo {author} {\bibfnamefont {G.~T.}\ \bibnamefont
  {Genov}}, \bibinfo {author} {\bibfnamefont {D.}~\bibnamefont {Schraft}},
  \bibinfo {author} {\bibfnamefont {N.~V.}\ \bibnamefont {Vitanov}}, \ and\
  \bibinfo {author} {\bibfnamefont {T.}~\bibnamefont {Halfmann}},\ }\bibinfo
  {title} {Arbitrarily accurate pulse sequences for robust dynamical
  decoupling},\ \href {\doibase 10.1103/PhysRevLett.118.133202} {\bibfield
  {journal} {\bibinfo  {journal} {Phys. Rev. Lett.}\ }\textbf {\bibinfo
  {volume} {118}},\ \bibinfo {pages} {133202} (\bibinfo {year}
  {2017})}\BibitemShut {NoStop}%
\bibitem [{\citenamefont {Torosov}\ and\ \citenamefont
  {Vitanov}(2019{\natexlab{c}})}]{PhysRevA.100.023410}%
  \BibitemOpen
  \bibfield  {author} {\bibinfo {author} {\bibfnamefont {B.~T.}\ \bibnamefont
  {Torosov}}\ and\ \bibinfo {author} {\bibfnamefont {N.~V.}\ \bibnamefont
  {Vitanov}},\ }\bibinfo {title} {Composite pulses with errant phases},\ \href
  {\doibase 10.1103/PhysRevA.100.023410} {\bibfield  {journal} {\bibinfo
  {journal} {Phys. Rev. A}\ }\textbf {\bibinfo {volume} {100}},\ \bibinfo
  {pages} {023410} (\bibinfo {year} {2019}{\natexlab{c}})}\BibitemShut
  {NoStop}%
\bibitem [{\citenamefont {Torosov}\ \emph
  {et~al.}(2020{\natexlab{a}})\citenamefont {Torosov}, \citenamefont {Ivanov},\
  and\ \citenamefont {Vitanov}}]{PhysRevA.102.013105}%
  \BibitemOpen
  \bibfield  {author} {\bibinfo {author} {\bibfnamefont {B.~T.}\ \bibnamefont
  {Torosov}}, \bibinfo {author} {\bibfnamefont {S.~S.}\ \bibnamefont {Ivanov}},
  \ and\ \bibinfo {author} {\bibfnamefont {N.~V.}\ \bibnamefont {Vitanov}},\
  }\bibinfo {title} {Narrowband and passband composite pulses for variable
  rotations},\ \href {\doibase 10.1103/PhysRevA.102.013105} {\bibfield
  {journal} {\bibinfo  {journal} {Phys. Rev. A}\ }\textbf {\bibinfo {volume}
  {102}},\ \bibinfo {pages} {013105} (\bibinfo {year}
  {2020}{\natexlab{a}})}\BibitemShut {NoStop}%
\bibitem [{\citenamefont {Genov}\ \emph {et~al.}(2011)\citenamefont {Genov},
  \citenamefont {Torosov},\ and\ \citenamefont {Vitanov}}]{PhysRevA.84.063413}%
  \BibitemOpen
  \bibfield  {author} {\bibinfo {author} {\bibfnamefont {G.~T.}\ \bibnamefont
  {Genov}}, \bibinfo {author} {\bibfnamefont {B.~T.}\ \bibnamefont {Torosov}},
  \ and\ \bibinfo {author} {\bibfnamefont {N.~V.}\ \bibnamefont {Vitanov}},\
  }\bibinfo {title} {Optimized control of multistate quantum systems by
  composite pulse sequences},\ \href {\doibase 10.1103/PhysRevA.84.063413}
  {\bibfield  {journal} {\bibinfo  {journal} {Phys. Rev. A}\ }\textbf {\bibinfo
  {volume} {84}},\ \bibinfo {pages} {063413} (\bibinfo {year}
  {2011})}\BibitemShut {NoStop}%
\bibitem [{\citenamefont {Randall}\ \emph {et~al.}(2018)\citenamefont
  {Randall}, \citenamefont {Lawrence}, \citenamefont {Webster}, \citenamefont
  {Weidt}, \citenamefont {Vitanov},\ and\ \citenamefont
  {Hensinger}}]{PhysRevA.98.043414}%
  \BibitemOpen
  \bibfield  {author} {\bibinfo {author} {\bibfnamefont {J.}~\bibnamefont
  {Randall}}, \bibinfo {author} {\bibfnamefont {A.~M.}\ \bibnamefont
  {Lawrence}}, \bibinfo {author} {\bibfnamefont {S.~C.}\ \bibnamefont
  {Webster}}, \bibinfo {author} {\bibfnamefont {S.}~\bibnamefont {Weidt}},
  \bibinfo {author} {\bibfnamefont {N.~V.}\ \bibnamefont {Vitanov}}, \ and\
  \bibinfo {author} {\bibfnamefont {W.~K.}\ \bibnamefont {Hensinger}},\
  }\bibinfo {title} {Generation of high-fidelity quantum control methods for
  multilevel systems},\ \href {\doibase 10.1103/PhysRevA.98.043414} {\bibfield
  {journal} {\bibinfo  {journal} {Phys. Rev. A}\ }\textbf {\bibinfo {volume}
  {98}},\ \bibinfo {pages} {043414} (\bibinfo {year} {2018})}\BibitemShut
  {NoStop}%
\bibitem [{\citenamefont {Greener}\ and\ \citenamefont
  {Suchowski}(2018)}]{Greener2018}%
  \BibitemOpen
  \bibfield  {author} {\bibinfo {author} {\bibfnamefont {H.}~\bibnamefont
  {Greener}}\ and\ \bibinfo {author} {\bibfnamefont {H.}~\bibnamefont
  {Suchowski}},\ }\bibinfo {title} {Composite pulses in {N}-level systems with
  {SU}(2) symmetry and their geometrical representation on the {M}ajorana
  sphere},\ \href {\doibase 10.1063/1.5013672} {\bibfield  {journal} {\bibinfo
  {journal} {The Journal of Chemical Physics}\ }\textbf {\bibinfo {volume}
  {148}},\ \bibinfo {pages} {074101} (\bibinfo {year} {2018})}\BibitemShut
  {NoStop}%
\bibitem [{\citenamefont {Torosov}\ and\ \citenamefont
  {Vitanov}(2020)}]{PhysRevResearch.2.043194}%
  \BibitemOpen
  \bibfield  {author} {\bibinfo {author} {\bibfnamefont {B.~T.}\ \bibnamefont
  {Torosov}}\ and\ \bibinfo {author} {\bibfnamefont {N.~V.}\ \bibnamefont
  {Vitanov}},\ }\bibinfo {title} {High-fidelity composite quantum gates for
  {R}aman qubits},\ \href {\doibase 10.1103/PhysRevResearch.2.043194}
  {\bibfield  {journal} {\bibinfo  {journal} {Phys. Rev. Research}\ }\textbf
  {\bibinfo {volume} {2}},\ \bibinfo {pages} {043194} (\bibinfo {year}
  {2020})}\BibitemShut {NoStop}%
\bibitem [{\citenamefont {Torosov}\ \emph
  {et~al.}(2020{\natexlab{b}})\citenamefont {Torosov}, \citenamefont
  {Drewsen},\ and\ \citenamefont {Vitanov}}]{PhysRevResearch.2.043235}%
  \BibitemOpen
  \bibfield  {author} {\bibinfo {author} {\bibfnamefont {B.~T.}\ \bibnamefont
  {Torosov}}, \bibinfo {author} {\bibfnamefont {M.}~\bibnamefont {Drewsen}}, \
  and\ \bibinfo {author} {\bibfnamefont {N.~V.}\ \bibnamefont {Vitanov}},\
  }\bibinfo {title} {Chiral resolution by composite {R}aman pulses},\ \href
  {\doibase 10.1103/PhysRevResearch.2.043235} {\bibfield  {journal} {\bibinfo
  {journal} {Phys. Rev. Research}\ }\textbf {\bibinfo {volume} {2}},\ \bibinfo
  {pages} {043235} (\bibinfo {year} {2020}{\natexlab{b}})}\BibitemShut
  {NoStop}%
\bibitem [{\citenamefont {Breuer}\ and\ \citenamefont
  {Petruccione}(2006)}]{Breuer2006}%
  \BibitemOpen
  \bibfield  {author} {\bibinfo {author} {\bibfnamefont {H.~P.}\ \bibnamefont
  {Breuer}}\ and\ \bibinfo {author} {\bibfnamefont {F.}~\bibnamefont
  {Petruccione}},\ }\href@noop {} {\emph {\bibinfo {title} {The Theory of Open
  Quantum Systems}}}\ (\bibinfo  {publisher} {Oxford University Press},\
  \bibinfo {year} {2006})\BibitemShut {NoStop}%
\bibitem [{\citenamefont {Omran}\ \emph {et~al.}(2019)\citenamefont {Omran},
  \citenamefont {Levine}, \citenamefont {Keesling}, \citenamefont {Semeghini},
  \citenamefont {Wang}, \citenamefont {Ebadi}, \citenamefont {Bernien},
  \citenamefont {Zibrov}, \citenamefont {Pichler}, \citenamefont {Choi},
  \citenamefont {Cui}, \citenamefont {Rossignolo}, \citenamefont {Rembold},
  \citenamefont {Montangero}, \citenamefont {Calarco}, \citenamefont {Endres},
  \citenamefont {Greiner}, \citenamefont {Vuleti{\'{c}}},\ and\ \citenamefont
  {Lukin}}]{Omran2019}%
  \BibitemOpen
  \bibfield  {author} {\bibinfo {author} {\bibfnamefont {A.}~\bibnamefont
  {Omran}}, \bibinfo {author} {\bibfnamefont {H.}~\bibnamefont {Levine}},
  \bibinfo {author} {\bibfnamefont {A.}~\bibnamefont {Keesling}}, \bibinfo
  {author} {\bibfnamefont {G.}~\bibnamefont {Semeghini}}, \bibinfo {author}
  {\bibfnamefont {T.~T.}\ \bibnamefont {Wang}}, \bibinfo {author}
  {\bibfnamefont {S.}~\bibnamefont {Ebadi}}, \bibinfo {author} {\bibfnamefont
  {H.}~\bibnamefont {Bernien}}, \bibinfo {author} {\bibfnamefont {A.~S.}\
  \bibnamefont {Zibrov}}, \bibinfo {author} {\bibfnamefont {H.}~\bibnamefont
  {Pichler}}, \bibinfo {author} {\bibfnamefont {S.}~\bibnamefont {Choi}},
  \bibinfo {author} {\bibfnamefont {J.}~\bibnamefont {Cui}}, \bibinfo {author}
  {\bibfnamefont {M.}~\bibnamefont {Rossignolo}}, \bibinfo {author}
  {\bibfnamefont {P.}~\bibnamefont {Rembold}}, \bibinfo {author} {\bibfnamefont
  {S.}~\bibnamefont {Montangero}}, \bibinfo {author} {\bibfnamefont
  {T.}~\bibnamefont {Calarco}}, \bibinfo {author} {\bibfnamefont
  {M.}~\bibnamefont {Endres}}, \bibinfo {author} {\bibfnamefont
  {M.}~\bibnamefont {Greiner}}, \bibinfo {author} {\bibfnamefont
  {V.}~\bibnamefont {Vuleti{\'{c}}}}, \ and\ \bibinfo {author} {\bibfnamefont
  {M.~D.}\ \bibnamefont {Lukin}},\ }\bibinfo {title} {Generation and
  manipulation of {S}chr\"{o}dinger cat states in {R}ydberg atom arrays},\
  \href {\doibase 10.1126/science.aax9743} {\bibfield  {journal} {\bibinfo
  {journal} {Science}\ }\textbf {\bibinfo {volume} {365}},\ \bibinfo {pages}
  {570} (\bibinfo {year} {2019})}\BibitemShut {NoStop}%
\bibitem [{\citenamefont {Madjarov}\ \emph {et~al.}(2020)\citenamefont
  {Madjarov}, \citenamefont {Covey}, \citenamefont {Shaw}, \citenamefont
  {Choi}, \citenamefont {Kale}, \citenamefont {Cooper}, \citenamefont
  {Pichler}, \citenamefont {Schkolnik}, \citenamefont {Williams},\ and\
  \citenamefont {Endres}}]{Madjarov2020}%
  \BibitemOpen
  \bibfield  {author} {\bibinfo {author} {\bibfnamefont {I.~S.}\ \bibnamefont
  {Madjarov}}, \bibinfo {author} {\bibfnamefont {J.~P.}\ \bibnamefont {Covey}},
  \bibinfo {author} {\bibfnamefont {A.~L.}\ \bibnamefont {Shaw}}, \bibinfo
  {author} {\bibfnamefont {J.}~\bibnamefont {Choi}}, \bibinfo {author}
  {\bibfnamefont {A.}~\bibnamefont {Kale}}, \bibinfo {author} {\bibfnamefont
  {A.}~\bibnamefont {Cooper}}, \bibinfo {author} {\bibfnamefont
  {H.}~\bibnamefont {Pichler}}, \bibinfo {author} {\bibfnamefont
  {V.}~\bibnamefont {Schkolnik}}, \bibinfo {author} {\bibfnamefont {J.~R.}\
  \bibnamefont {Williams}}, \ and\ \bibinfo {author} {\bibfnamefont
  {M.}~\bibnamefont {Endres}},\ }\bibinfo {title} {High-fidelity entanglement
  and detection of alkaline-earth {R}ydberg atoms},\ \href {\doibase
  10.1038/s41567-020-0903-z} {\bibfield  {journal} {\bibinfo  {journal} {Nature
  Physics}\ }\textbf {\bibinfo {volume} {16}},\ \bibinfo {pages} {857}
  (\bibinfo {year} {2020})}\BibitemShut {NoStop}%
\bibitem [{\citenamefont {Barredo}\ \emph {et~al.}(2015)\citenamefont
  {Barredo}, \citenamefont {Labuhn}, \citenamefont {Ravets}, \citenamefont
  {Lahaye}, \citenamefont {Browaeys},\ and\ \citenamefont
  {Adams}}]{PhysRevLett.114.113002}%
  \BibitemOpen
  \bibfield  {author} {\bibinfo {author} {\bibfnamefont {D.}~\bibnamefont
  {Barredo}}, \bibinfo {author} {\bibfnamefont {H.}~\bibnamefont {Labuhn}},
  \bibinfo {author} {\bibfnamefont {S.}~\bibnamefont {Ravets}}, \bibinfo
  {author} {\bibfnamefont {T.}~\bibnamefont {Lahaye}}, \bibinfo {author}
  {\bibfnamefont {A.}~\bibnamefont {Browaeys}}, \ and\ \bibinfo {author}
  {\bibfnamefont {C.~S.}\ \bibnamefont {Adams}},\ }\bibinfo {title} {Coherent
  excitation transfer in a spin chain of three rydberg atoms},\ \href {\doibase
  10.1103/PhysRevLett.114.113002} {\bibfield  {journal} {\bibinfo  {journal}
  {Phys. Rev. Lett.}\ }\textbf {\bibinfo {volume} {114}},\ \bibinfo {pages}
  {113002} (\bibinfo {year} {2015})}\BibitemShut {NoStop}%
\bibitem [{\citenamefont {Scully}\ and\ \citenamefont
  {Zubairy}(1997)}]{scully97}%
  \BibitemOpen
  \bibfield  {author} {\bibinfo {author} {\bibfnamefont {M.~O.}\ \bibnamefont
  {Scully}}\ and\ \bibinfo {author} {\bibfnamefont {M.~S.}\ \bibnamefont
  {Zubairy}},\ }\href {\doibase 10.1017/cbo9780511813993} {\emph {\bibinfo
  {title} {Quantum Optics}}}\ (\bibinfo  {publisher} {Cambridge University
  Press},\ \bibinfo {year} {1997})\BibitemShut {NoStop}%
\bibitem [{\citenamefont {Leibfried}\ \emph {et~al.}(2003)\citenamefont
  {Leibfried}, \citenamefont {Blatt}, \citenamefont {Monroe},\ and\
  \citenamefont {Wineland}}]{RevModPhys.75.281}%
  \BibitemOpen
  \bibfield  {author} {\bibinfo {author} {\bibfnamefont {D.}~\bibnamefont
  {Leibfried}}, \bibinfo {author} {\bibfnamefont {R.}~\bibnamefont {Blatt}},
  \bibinfo {author} {\bibfnamefont {C.}~\bibnamefont {Monroe}}, \ and\ \bibinfo
  {author} {\bibfnamefont {D.}~\bibnamefont {Wineland}},\ }\bibinfo {title}
  {Quantum dynamics of single trapped ions},\ \href {\doibase
  10.1103/RevModPhys.75.281} {\bibfield  {journal} {\bibinfo  {journal} {Rev.
  Mod. Phys.}\ }\textbf {\bibinfo {volume} {75}},\ \bibinfo {pages} {281}
  (\bibinfo {year} {2003})}\BibitemShut {NoStop}%
\bibitem [{\citenamefont {Xiang}\ \emph {et~al.}(2013)\citenamefont {Xiang},
  \citenamefont {Ashhab}, \citenamefont {You},\ and\ \citenamefont
  {Nori}}]{RevModPhys.85.623}%
  \BibitemOpen
  \bibfield  {author} {\bibinfo {author} {\bibfnamefont {Z.-L.}\ \bibnamefont
  {Xiang}}, \bibinfo {author} {\bibfnamefont {S.}~\bibnamefont {Ashhab}},
  \bibinfo {author} {\bibfnamefont {J.~Q.}\ \bibnamefont {You}}, \ and\
  \bibinfo {author} {\bibfnamefont {F.}~\bibnamefont {Nori}},\ }\bibinfo
  {title} {Hybrid quantum circuits: Superconducting circuits interacting with
  other quantum systems},\ \href {\doibase 10.1103/RevModPhys.85.623}
  {\bibfield  {journal} {\bibinfo  {journal} {Rev. Mod. Phys.}\ }\textbf
  {\bibinfo {volume} {85}},\ \bibinfo {pages} {623} (\bibinfo {year}
  {2013})}\BibitemShut {NoStop}%
\bibitem [{\citenamefont {Pedersen}\ \emph {et~al.}(2007)\citenamefont
  {Pedersen}, \citenamefont {M{\o}ller},\ and\ \citenamefont
  {M{\o}lmer}}]{Pedersen2007}%
  \BibitemOpen
  \bibfield  {author} {\bibinfo {author} {\bibfnamefont {L.~H.}\ \bibnamefont
  {Pedersen}}, \bibinfo {author} {\bibfnamefont {N.~M.}\ \bibnamefont
  {M{\o}ller}}, \ and\ \bibinfo {author} {\bibfnamefont {K.}~\bibnamefont
  {M{\o}lmer}},\ }\bibinfo {title} {Fidelity of quantum operations},\ \href
  {\doibase 10.1016/j.physleta.2007.02.069} {\bibfield  {journal} {\bibinfo
  {journal} {Physics Letters A}\ }\textbf {\bibinfo {volume} {367}},\ \bibinfo
  {pages} {47} (\bibinfo {year} {2007})}\BibitemShut {NoStop}%
\bibitem [{\citenamefont {Ghosh}\ and\ \citenamefont {Geller}(2010)}]{Ghosh10}%
  \BibitemOpen
  \bibfield  {author} {\bibinfo {author} {\bibfnamefont {J.}~\bibnamefont
  {Ghosh}}\ and\ \bibinfo {author} {\bibfnamefont {M.~R.}\ \bibnamefont
  {Geller}},\ }\bibinfo {title} {Controlled-{NOT} gate with weakly coupled
  qubits: {D}ependence of fidelity on the form of interaction},\ \href
  {\doibase 10.1103/PhysRevA.81.052340} {\bibfield  {journal} {\bibinfo
  {journal} {Phys. Rev. A}\ }\textbf {\bibinfo {volume} {81}},\ \bibinfo
  {pages} {052340} (\bibinfo {year} {2010})}\BibitemShut {NoStop}%
\bibitem [{\citenamefont {Bose}(2003)}]{PhysRevLett.91.207901}%
  \BibitemOpen
  \bibfield  {author} {\bibinfo {author} {\bibfnamefont {S.}~\bibnamefont
  {Bose}},\ }\bibinfo {title} {Quantum communication through an unmodulated
  spin chain},\ \href {\doibase 10.1103/PhysRevLett.91.207901} {\bibfield
  {journal} {\bibinfo  {journal} {Phys. Rev. Lett.}\ }\textbf {\bibinfo
  {volume} {91}},\ \bibinfo {pages} {207901} (\bibinfo {year}
  {2003})}\BibitemShut {NoStop}%
\bibitem [{\citenamefont {Christandl}\ \emph {et~al.}(2004)\citenamefont
  {Christandl}, \citenamefont {Datta}, \citenamefont {Ekert},\ and\
  \citenamefont {Landahl}}]{PhysRevLett.92.187902}%
  \BibitemOpen
  \bibfield  {author} {\bibinfo {author} {\bibfnamefont {M.}~\bibnamefont
  {Christandl}}, \bibinfo {author} {\bibfnamefont {N.}~\bibnamefont {Datta}},
  \bibinfo {author} {\bibfnamefont {A.}~\bibnamefont {Ekert}}, \ and\ \bibinfo
  {author} {\bibfnamefont {A.~J.}\ \bibnamefont {Landahl}},\ }\bibinfo {title}
  {Perfect state transfer in quantum spin networks},\ \href {\doibase
  10.1103/PhysRevLett.92.187902} {\bibfield  {journal} {\bibinfo  {journal}
  {Phys. Rev. Lett.}\ }\textbf {\bibinfo {volume} {92}},\ \bibinfo {pages}
  {187902} (\bibinfo {year} {2004})}\BibitemShut {NoStop}%
\bibitem [{\citenamefont {Bayat}\ and\ \citenamefont
  {Bose}(2010)}]{PhysRevA.81.012304}%
  \BibitemOpen
  \bibfield  {author} {\bibinfo {author} {\bibfnamefont {A.}~\bibnamefont
  {Bayat}}\ and\ \bibinfo {author} {\bibfnamefont {S.}~\bibnamefont {Bose}},\
  }\bibinfo {title} {Information-transferring ability of the different phases
  of a finite {XXZ} spin chain},\ \href {\doibase 10.1103/PhysRevA.81.012304}
  {\bibfield  {journal} {\bibinfo  {journal} {Phys. Rev. A}\ }\textbf {\bibinfo
  {volume} {81}},\ \bibinfo {pages} {012304} (\bibinfo {year}
  {2010})}\BibitemShut {NoStop}%
\bibitem [{\citenamefont {Shi}\ \emph {et~al.}(2015)\citenamefont {Shi},
  \citenamefont {Zhao},\ and\ \citenamefont {Yi}}]{PhysRevA.91.032301}%
  \BibitemOpen
  \bibfield  {author} {\bibinfo {author} {\bibfnamefont {Z.~C.}\ \bibnamefont
  {Shi}}, \bibinfo {author} {\bibfnamefont {X.~L.}\ \bibnamefont {Zhao}}, \
  and\ \bibinfo {author} {\bibfnamefont {X.~X.}\ \bibnamefont {Yi}},\ }\bibinfo
  {title} {Robust state transfer with high fidelity in spin-1/2 chains by
  {L}yapunov control},\ \href {\doibase 10.1103/PhysRevA.91.032301} {\bibfield
  {journal} {\bibinfo  {journal} {Phys. Rev. A}\ }\textbf {\bibinfo {volume}
  {91}},\ \bibinfo {pages} {032301} (\bibinfo {year} {2015})}\BibitemShut
  {NoStop}%
\bibitem [{\citenamefont {Yousefjani}\ and\ \citenamefont
  {Bayat}(2020{\natexlab{a}})}]{PhysRevA.102.012418}%
  \BibitemOpen
  \bibfield  {author} {\bibinfo {author} {\bibfnamefont {R.}~\bibnamefont
  {Yousefjani}}\ and\ \bibinfo {author} {\bibfnamefont {A.}~\bibnamefont
  {Bayat}},\ }\bibinfo {title} {Simultaneous multiple-user quantum
  communication across a spin-chain channel},\ \href {\doibase
  10.1103/PhysRevA.102.012418} {\bibfield  {journal} {\bibinfo  {journal}
  {Phys. Rev. A}\ }\textbf {\bibinfo {volume} {102}},\ \bibinfo {pages}
  {012418} (\bibinfo {year} {2020}{\natexlab{a}})}\BibitemShut {NoStop}%
\bibitem [{\citenamefont {Yousefjani}\ and\ \citenamefont
  {Bayat}(2020{\natexlab{b}})}]{Yousefjani20}%
  \BibitemOpen
  \bibfield  {author} {\bibinfo {author} {\bibfnamefont {R.}~\bibnamefont
  {Yousefjani}}\ and\ \bibinfo {author} {\bibfnamefont {A.}~\bibnamefont
  {Bayat}},\ }\bibinfo {title} {Parallel entangling gate operations and two-way
  quantum communication in spin chains},\ \href
  {https://arxiv.org/abs/2008.12771} {\bibfield  {journal} {\bibinfo  {journal}
  {arXiv: 2008.12771}\ } (\bibinfo {year} {2020}{\natexlab{b}})}\BibitemShut
  {NoStop}%
\bibitem [{\citenamefont {Yang}\ \emph {et~al.}(2010)\citenamefont {Yang},
  \citenamefont {Xu}, \citenamefont {Feng},\ and\ \citenamefont
  {Du}}]{Yang2010}%
  \BibitemOpen
  \bibfield  {author} {\bibinfo {author} {\bibfnamefont {W.}~\bibnamefont
  {Yang}}, \bibinfo {author} {\bibfnamefont {Z.}~\bibnamefont {Xu}}, \bibinfo
  {author} {\bibfnamefont {M.}~\bibnamefont {Feng}}, \ and\ \bibinfo {author}
  {\bibfnamefont {J.}~\bibnamefont {Du}},\ }\bibinfo {title} {Entanglement of
  separate nitrogen-vacancy centers coupled to a whispering-gallery mode
  cavity},\ \href {\doibase 10.1088/1367-2630/12/11/113039} {\bibfield
  {journal} {\bibinfo  {journal} {New Journal of Physics}\ }\textbf {\bibinfo
  {volume} {12}},\ \bibinfo {pages} {113039} (\bibinfo {year}
  {2010})}\BibitemShut {NoStop}%
\bibitem [{\citenamefont {Sterk}\ \emph {et~al.}(2012)\citenamefont {Sterk},
  \citenamefont {Luo}, \citenamefont {Manning}, \citenamefont {Maunz},\ and\
  \citenamefont {Monroe}}]{PhysRevA.85.062308}%
  \BibitemOpen
  \bibfield  {author} {\bibinfo {author} {\bibfnamefont {J.~D.}\ \bibnamefont
  {Sterk}}, \bibinfo {author} {\bibfnamefont {L.}~\bibnamefont {Luo}}, \bibinfo
  {author} {\bibfnamefont {T.~A.}\ \bibnamefont {Manning}}, \bibinfo {author}
  {\bibfnamefont {P.}~\bibnamefont {Maunz}}, \ and\ \bibinfo {author}
  {\bibfnamefont {C.}~\bibnamefont {Monroe}},\ }\bibinfo {title} {Photon
  collection from a trapped ion-cavity system},\ \href {\doibase
  10.1103/PhysRevA.85.062308} {\bibfield  {journal} {\bibinfo  {journal} {Phys.
  Rev. A}\ }\textbf {\bibinfo {volume} {85}},\ \bibinfo {pages} {062308}
  (\bibinfo {year} {2012})}\BibitemShut {NoStop}%
\bibitem [{\citenamefont {Huang}\ \emph {et~al.}(2017)\citenamefont {Huang},
  \citenamefont {Kang}, \citenamefont {Chen}, \citenamefont {Wu}, \citenamefont
  {Song},\ and\ \citenamefont {Xia}}]{PhysRevA.96.022314}%
  \BibitemOpen
  \bibfield  {author} {\bibinfo {author} {\bibfnamefont {B.-H.}\ \bibnamefont
  {Huang}}, \bibinfo {author} {\bibfnamefont {Y.-H.}\ \bibnamefont {Kang}},
  \bibinfo {author} {\bibfnamefont {Y.-H.}\ \bibnamefont {Chen}}, \bibinfo
  {author} {\bibfnamefont {Q.-C.}\ \bibnamefont {Wu}}, \bibinfo {author}
  {\bibfnamefont {J.}~\bibnamefont {Song}}, \ and\ \bibinfo {author}
  {\bibfnamefont {Y.}~\bibnamefont {Xia}},\ }\bibinfo {title} {Fast quantum
  state engineering via universal {SU}(2) transformation},\ \href {\doibase
  10.1103/PhysRevA.96.022314} {\bibfield  {journal} {\bibinfo  {journal} {Phys.
  Rev. A}\ }\textbf {\bibinfo {volume} {96}},\ \bibinfo {pages} {022314}
  (\bibinfo {year} {2017})}\BibitemShut {NoStop}%
\bibitem [{\citenamefont {Wollenhaupt}\ \emph {et~al.}(2005)\citenamefont
  {Wollenhaupt}, \citenamefont {Engel},\ and\ \citenamefont
  {Baumert}}]{Wollenhaupt2005}%
  \BibitemOpen
  \bibfield  {author} {\bibinfo {author} {\bibfnamefont {M.}~\bibnamefont
  {Wollenhaupt}}, \bibinfo {author} {\bibfnamefont {V.}~\bibnamefont {Engel}},
  \ and\ \bibinfo {author} {\bibfnamefont {T.}~\bibnamefont {Baumert}},\
  }\bibinfo {title} {Femtosecond laser photoelectron spectroscopy on atoms and
  small molecules: Prototype studies in quantum control},\ \href {\doibase
  10.1146/annurev.physchem.56.092503.141315} {\bibfield  {journal} {\bibinfo
  {journal} {Annu. Rev. Phys. Chem.}\ }\textbf {\bibinfo {volume} {56}},\
  \bibinfo {pages} {25} (\bibinfo {year} {2005})}\BibitemShut {NoStop}%
\bibitem [{\citenamefont {Cho}\ \emph {et~al.}(2008)\citenamefont {Cho},
  \citenamefont {Angelakis},\ and\ \citenamefont {Bose}}]{cho2008}%
  \BibitemOpen
  \bibfield  {author} {\bibinfo {author} {\bibfnamefont {J.}~\bibnamefont
  {Cho}}, \bibinfo {author} {\bibfnamefont {D.~G.}\ \bibnamefont {Angelakis}},
  \ and\ \bibinfo {author} {\bibfnamefont {S.}~\bibnamefont {Bose}},\ }\bibinfo
  {title} {Fractional quantum {H}all state in coupled cavities},\ \href
  {\doibase 10.1103/PhysRevLett.101.246809} {\bibfield  {journal} {\bibinfo
  {journal} {Phys. Rev. Lett.}\ }\textbf {\bibinfo {volume} {101}},\ \bibinfo
  {pages} {246809} (\bibinfo {year} {2008})}\BibitemShut {NoStop}%
\bibitem [{\citenamefont {Chen}\ \emph {et~al.}(2010)\citenamefont {Chen},
  \citenamefont {Zhou}, \citenamefont {Zhou}, \citenamefont {Zhou},\ and\
  \citenamefont {Guo}}]{chen2010}%
  \BibitemOpen
  \bibfield  {author} {\bibinfo {author} {\bibfnamefont {Z.-X.}\ \bibnamefont
  {Chen}}, \bibinfo {author} {\bibfnamefont {Z.-W.}\ \bibnamefont {Zhou}},
  \bibinfo {author} {\bibfnamefont {X.}~\bibnamefont {Zhou}}, \bibinfo {author}
  {\bibfnamefont {X.-F.}\ \bibnamefont {Zhou}}, \ and\ \bibinfo {author}
  {\bibfnamefont {G.-C.}\ \bibnamefont {Guo}},\ }\bibinfo {title} {Quantum
  simulation of {H}eisenberg spin chains with next-nearest-neighbor
  interactions in coupled cavities},\ \href {\doibase
  10.1103/PhysRevA.81.022303} {\bibfield  {journal} {\bibinfo  {journal} {Phys.
  Rev. A}\ }\textbf {\bibinfo {volume} {81}},\ \bibinfo {pages} {022303}
  (\bibinfo {year} {2010})}\BibitemShut {NoStop}%
\bibitem [{\citenamefont {Keating}\ \emph {et~al.}(2016)\citenamefont
  {Keating}, \citenamefont {Baldwin}, \citenamefont {Jau}, \citenamefont {Lee},
  \citenamefont {Biedermann},\ and\ \citenamefont {Deutsch}}]{keating2016}%
  \BibitemOpen
  \bibfield  {author} {\bibinfo {author} {\bibfnamefont {T.}~\bibnamefont
  {Keating}}, \bibinfo {author} {\bibfnamefont {C.~H.}\ \bibnamefont
  {Baldwin}}, \bibinfo {author} {\bibfnamefont {Y.-Y.}\ \bibnamefont {Jau}},
  \bibinfo {author} {\bibfnamefont {J.}~\bibnamefont {Lee}}, \bibinfo {author}
  {\bibfnamefont {G.~W.}\ \bibnamefont {Biedermann}}, \ and\ \bibinfo {author}
  {\bibfnamefont {I.~H.}\ \bibnamefont {Deutsch}},\ }\bibinfo {title}
  {Arbitrary {D}icke-state control of symmetric {R}ydberg ensembles},\ \href
  {\doibase 10.1103/PhysRevLett.117.213601} {\bibfield  {journal} {\bibinfo
  {journal} {Phys. Rev. Lett.}\ }\textbf {\bibinfo {volume} {117}},\ \bibinfo
  {pages} {213601} (\bibinfo {year} {2016})}\BibitemShut {NoStop}%
\bibitem [{\citenamefont {Zhang}\ \emph {et~al.}(2017)\citenamefont {Zhang},
  \citenamefont {Zhou}, \citenamefont {Zhou}, \citenamefont {Guo},\ and\
  \citenamefont {Zhou}}]{zhang17}%
  \BibitemOpen
  \bibfield  {author} {\bibinfo {author} {\bibfnamefont {Y.-C.}\ \bibnamefont
  {Zhang}}, \bibinfo {author} {\bibfnamefont {X.-F.}\ \bibnamefont {Zhou}},
  \bibinfo {author} {\bibfnamefont {X.}~\bibnamefont {Zhou}}, \bibinfo {author}
  {\bibfnamefont {G.-C.}\ \bibnamefont {Guo}}, \ and\ \bibinfo {author}
  {\bibfnamefont {Z.-W.}\ \bibnamefont {Zhou}},\ }\bibinfo {title}
  {Cavity-assisted single-mode and two-mode spin-squeezed states via
  phase-locked atom-photon coupling},\ \href {\doibase
  10.1103/PhysRevLett.118.083604} {\bibfield  {journal} {\bibinfo  {journal}
  {Phys. Rev. Lett.}\ }\textbf {\bibinfo {volume} {118}},\ \bibinfo {pages}
  {083604} (\bibinfo {year} {2017})}\BibitemShut {NoStop}%
\bibitem [{\citenamefont {Shi}\ and\ \citenamefont {Kennedy}(2017)}]{shi17}%
  \BibitemOpen
  \bibfield  {author} {\bibinfo {author} {\bibfnamefont {X.-F.}\ \bibnamefont
  {Shi}}\ and\ \bibinfo {author} {\bibfnamefont {T.~A.~B.}\ \bibnamefont
  {Kennedy}},\ }\bibinfo {title} {Annulled van der {W}aals interaction and fast
  {R}ydberg quantum gates},\ \href {\doibase 10.1103/PhysRevA.95.043429}
  {\bibfield  {journal} {\bibinfo  {journal} {Phys. Rev. A}\ }\textbf {\bibinfo
  {volume} {95}},\ \bibinfo {pages} {043429} (\bibinfo {year}
  {2017})}\BibitemShut {NoStop}%
\bibitem [{\citenamefont {Li}\ \emph {et~al.}(2018)\citenamefont {Li},
  \citenamefont {Mart\'{\i}nez-Cerc\'os}, \citenamefont {Mart\'{\i}nez-Garaot},
  \citenamefont {Chen},\ and\ \citenamefont {Muga}}]{PhysRevA.97.013830}%
  \BibitemOpen
  \bibfield  {author} {\bibinfo {author} {\bibfnamefont {Y.-C.}\ \bibnamefont
  {Li}}, \bibinfo {author} {\bibfnamefont {D.}~\bibnamefont
  {Mart\'{\i}nez-Cerc\'os}}, \bibinfo {author} {\bibfnamefont {S.}~\bibnamefont
  {Mart\'{\i}nez-Garaot}}, \bibinfo {author} {\bibfnamefont {X.}~\bibnamefont
  {Chen}}, \ and\ \bibinfo {author} {\bibfnamefont {J.~G.}\ \bibnamefont
  {Muga}},\ }\bibinfo {title} {Hamiltonian design to prepare arbitrary states
  of four-level systems},\ \href {\doibase 10.1103/PhysRevA.97.013830}
  {\bibfield  {journal} {\bibinfo  {journal} {Phys. Rev. A}\ }\textbf {\bibinfo
  {volume} {97}},\ \bibinfo {pages} {013830} (\bibinfo {year}
  {2018})}\BibitemShut {NoStop}%
\end{thebibliography}%

\end{document}